\documentclass[letter,journal]{IEEEtran}
\usepackage{amsmath} 
\usepackage{amssymb}
\usepackage{amsfonts}
\usepackage{amsthm}
\allowdisplaybreaks[4]
\usepackage{algorithmic}
\usepackage{algorithm}
\usepackage{gensymb}
\usepackage{array}
\usepackage{booktabs}
\usepackage[caption=false,font=normalsize,labelfont=sf,textfont=sf]{subfig}
\usepackage{textcomp}
\usepackage{color}
\usepackage{stfloats}
\usepackage{caption}
\usepackage{url}
\usepackage{verbatim}
\usepackage{graphicx}
\usepackage{diagbox}
\usepackage{epstopdf}
\usepackage{cite}

\newtheorem{lemma}{Lemma}
\newtheorem{prop}{Proposition}
\newtheorem{remark}{Remark}

\begin{document}

\title{Rate-Splitting Multiple Access for Transmissive Reconfigurable Intelligent Surface Transceiver Empowered ISAC Systems}

\author{Ziwei~Liu,~Wen~Chen,~\IEEEmembership{Senior~Member,~IEEE,}~Qingqing~Wu,~\IEEEmembership{Senior~Member,~IEEE,}~Jinhong~Yuan,~\IEEEmembership{Fellow,~IEEE},~Shanshan~Zhang,~Zhendong~Li,~and~Jun Li,~\IEEEmembership{Senior~Member,~IEEE}
	\thanks{(Corresponding author: Wen Chen.)}
	\thanks{Z. Liu, W. Chen, Q. Wu, S. Zhang, and Z. Li are with the Department of Electronic Engineering, Shanghai Jiao Tong University, Shanghai 200240, China (e-mail: ziweiliu@sjtu.edu.cn; wenchen@sjtu.edu.cn; qingqingwu@sjtu.edu.cn; shansz@sjtu.edu.cn;  lizhendong@sjtu.edu.cn).}
\thanks{J. Yuan is with the School of Electrical Engineering and Telecommunications, University of New South Wales, Sydney, NSW 2052, Australia (e-mail: j.yuan@unsw.edu.au). }	
\thanks{Jun Li is with the School of Electronic and Optical Engineering, Nanjing University of Science Technology, Nanjing 210094, China (e-mail: jun.li@njust.edu.cn).}

}

%

\maketitle

\begin{abstract}
In this paper, a novel transmissive reconfigurable intelligent surface (TRIS) transceiver empowered integrated sensing and communications (ISAC) system is proposed for future multi-demand terminals. To address interference management, we implement rate-splitting multiple access (RSMA), where the common stream is independently designed for the sensing service. We introduce the sensing quality of service (QoS) criteria based on this structure and construct an optimization problem with the sensing QoS criteria as the objective function to optimize the sensing stream precoding matrix and the communication stream precoding matrix. Due to the coupling of optimization variables, the formulated problem is a non-convex optimization problem that cannot be solved directly. To tackle the above-mentioned challenging problem, alternating optimization (AO) is utilized to decouple the optimization variables. Specifically, the problem is decoupled into three subproblems about the sensing stream precoding matrix, the communication stream precoding matrix, and the auxiliary variables, which is solved alternatively through AO until the convergence is reached. For solving the problem, successive convex approximation (SCA) is applied to deal with the sum-rate threshold constraints on communications, and difference-of-convex (DC) programming is utilized to solve rank-one non-convex constraints. Numerical simulation results verify the superiority of the proposed scheme in terms of improving the communication and sensing QoS.
\end{abstract}

\section{Introduction}
\IEEEPARstart{W}{ireless} communications have come a long way since its inception, enabling us to connect and communicate seamlessly across vast distances. Moving forward, the field of wireless communications is poised to make significant advances that will revolutionize the way we interact, exchange information, and experience the world\cite{Jiang2021}. While the future of wireless communications hold great promise, there are also potential challenges and problems that need to be addressed. One major challenge is the efficient utilization of limited resources, such as bandwidth, power, and computational capabilities. Another pressing challenge lies in the improvement of the wireless environment.

To tackle the problem of efficient utilization of resources, a solution for interaction with the environment to obtain information and allocate resources has come to the forefront, which is known as integrated sensing and communications (ISAC)\cite{Liu2022}. By continuously monitoring and analyzing factors such as network congestion, signal strength, and user demands, the system can dynamically allocate resources to optimize performance and ensure efficient utilization of available wireless resource. Previous works provide some algorithms for allocating resources\cite{Li2023,Li2022,Wang2017}, which consider communication sum-rate as the optimization metrics for the system, but in future communications network, the requirement of terminals will not be satisfied with communication, but with interaction with the wireless environment. Thus ISAC offers a new way of thinking. Due to the sharing of wireless resources, ISAC will provide an integral gain, and mutual assistance in communications and sensing will thrive their cooperative gains\cite{Liu2021}. The multi-user and multi-target dual-function radar-communication waveform design has been investigated in \cite{Du2023}, and the results illustrated the trade-off and coordination between sensing and communication. Moreover, the sensing and communication problems under vehicular communications network is investigated and performance enhancement of communications and sensing by developing functional ISAC is revealed in\cite{Cheng2022}. However, there is still the problem of mutual interference between communications and sensing, which has not been effectively addressed.

In the face of complex scenarios where communications and sensing coexist, the wireless environment is more complex and volatile, and interference management is an issue worth focusing on. By exploring advanced signal processing techniques, spectrum management strategies, and interference mitigation approaches, we can mitigate the effects of interference and enhance the reliability of wireless systems. Rate splitting multiple access (RSMA), as a general non-orthogonal access method, provides better interference management\cite{9832611,9831440}, which can achieve higher system gains than space division multiple access (SDMA) and non-orthogonal multiple access (NOMA)\cite{Mao2018}. Therefore, due to its excellent interference management capabilities, RSMA was adopted for the ISAC investigation\cite{Yin2022}. The RSMA-assisted dual-functional radar-communication (DFRC) beamforming was investigated in\cite{Xu2021,Yin2022a}, and the results demonstrated this RSMA scheme without radar sequences achieved better performance than the schemes using orthogonal resources. Moreover, RSMA maintains a larger communication and sensing trade-off than the conventional access schemes and can detect multiple targets with a high detection accuracy\cite{chen2023ratesplitting}. However, existing work does not separately consider and design the common stream of RSMA, which is not distinguished from the private stream. 

At the same time, there has been significant research conducted on reconfigurable intelligent surfaces (RIS) as an interference mitigation technique, owing to its exceptional capability to manipulate electromagnetic waves with unprecedented levels of flexibility and precision.\cite{Zhong2022,Sikri2022,Zhu2023a}. Reflective RIS improves the wireless environment by reconfiguring the channel, which is known as channel regulation. Due to the interference mitigation ability, RIS has emerged great research interest in ISAC. In \cite{Jiang2022,Zhu2023}, the RIS-aided ISAC system is investigated, and the results show that the RIS enhances the radar signal-to-noise ratio (SNR) significantly. In \cite{Luo2023}, the joint beamforming and reflection design under RIS-aided ISAC network has been investigated, and the results illustrate the advantages of deploying RIS in ISAC systems. In \cite{Sankar2022}, R.S. Prasobh Sankar {\it et~al.} investigate beamforming in hybrid RIS-assisted ISAC systems, and demonstrate that the performance of the hybrid RIS-assisted ISAC system is significantly better than that of passive RIS-assisted ISAC systems. In contrast, transmissive RIS (TRIS) serves as information regulation, where beam information is loaded onto the surface through time modulation array (TMA)\cite{he}, which in turn realizes the directional function of the beam. Due to the structural characteristics of TRIS, it avoids interference from reflected waves and occlusion from the feed source, and higher system gains can be realized in a low cost and low energy consumption way\cite{lizhendong}. In addition, TRIS-empowered RSMA networks were studied and the results showed that the combination of the two can achieve higher energy efficiency (EE) and spectral efficiency (SE) compared to the multi-antenna scheme\cite{Liu2023}.
 
Overall, the proliferation of future user demands and the increasing complexity of wireless environments have sparked research into the rational design of ISAC networks. Rational allocation of resources and management of interference in ISAC network is an issue of concern. To the best of our knowledge, the TRIS-empowered ISAC network is barely studied. In this paper, we adopt TRIS to empower ISAC network and provide a low-cost and low energy consumption architecture, in which RSMA is employed to manage the interference between communications and sensing. Based on the scenario requirements, an optimization problem with maximizing the sensing QoS as the objective function is constructed and resources are reasonably allocated to achieve coordination between communications and sensing.  The main contributions of this paper are as follows: 
\begin{itemize}
\item[$\bullet$] We propose a novel TRIS transceiver framework in the ISAC network. This architecture utilizes TMA to enable simultaneous multistream communications and sensing. Besides, the common stream of RSMA is independently designed and utilized to satisfy the different sensing demands of the terminals while ensuring the communication quality of the terminals. Based on these settings, the QoS criteria of sensing have been given, such as gain for detection, fisher information matrix (FIM) for localization and the posteriori FIM for tracking. Since the FIM with respect to the time delay is not derivable, in this paper we extend the received signal and transform the discrete received signal into a continuous one and derive an expression for the FIM with respect to the time delay, which has not been considered in previous works.
\item[$\bullet$] In response to the proposed network model, beamforming for ISAC is the problem to be solved. To solve this problem, we construct an optimization problem with sensing QoS as the objective function and communications QoS as the constraints, and then propose an alternating optimization (AO) algorithm to divide the constructed problem into three sub-problems. Specifically, in the first subproblem, the sensing stream precoding matrix is the first obtained based on the successive convex approximations (SCA) and the semidefinite relaxation (SDR) technique. In the second subproblem, difference-of-convex (DC) programming and semidefinite programming (SDP) are utilized  to solve the rank-one constraints and obtain the communication stream precoding matrices. In the third subproblem, we solve a linear programming (LP) problem to obtain the common stream rate vector. In addition, for the accuracy of the tracking, we introduce the extended Kalman filter (EKF) as an outer loop. Finally, the three subproblems are alternately iterated until the overall problem converges.
\item[$\bullet$] We evaluate the performance of the proposed ISAC scheme, and numerical simulation results show that the sensing QoS in the proposed architecture outperforms the benchmarks, which confirms the superiority of the proposed architecture. In addition, with the increase in the number of TRIS elements, the detection probability increases, the localization error decreases, and the beamwidth of the beampattern at the specific target angle decreases.
\end{itemize}

\emph{Notations}: Scalars are denoted by lower-case letters, while vectors and matrices are represented by bold lower-case letters and bold upper-case letters, respectively. $|x|$ denotes the absolute value of a complex-valued scalar $x$, ${x^ * }$ denotes the conjugate operation, and $\left\| \bf x \right\|$ denotes the Euclidean norm of a complex-valued vector $\bf x$. For a square matrix ${\bf{X}}$, ${\rm{tr}}\left( {\bf{X}} \right)$, ${\rm{rank}}\left( {\bf{X}} \right)$, ${{\bf{X}}^H}$, ${\left[ {\bf{X}} \right]_{m,n}}$ and $\left\| {\bf{X}} \right\|$ denote its trace, rank, conjugate transpose, ${m,n}$-th entry, and matrix norm, respectively. ${\bf{X}} \succeq 0$ represents that ${\bf{X}}$ is a positive semidefinite matrix. In addition, ${\mathbb{C}^{M \times N}}$ denotes the space of ${M \times N}$ complex matrices. $j$ denotes the imaginary element, i.e., $j^2 = -1$. ${\mathop{\rm Re}\nolimits} \left\{  \cdot  \right\}$ means the real part operation. The distribution of a circularly symmetric complex Gaussian (CSCG) random vector with mean $\mu $ and covariance matrix $\bf{C}$ is denoted by ${\cal C}{\cal N}\left( {{\mu},\bf{C}} \right)$ and $ \sim $ stands for ‘distributed as’. ${\bf{A}} \otimes {\bf{B}}$ represents the Kronecker product of matrices ${\bf{A}}$ and ${\bf{B}}$.

\section{System model}
In this paper, we consider an ISAC system based on a TRIS transceiver, where the single station is shared by the radar system and the communications system serving $K$ terminals. All terminals have both communication and sensing demands\footnote{For analytical convenience, three terminals are set up, all of which have communication demands, while at the same time they have detection, localization and tracking demands, respectively. In this paper, the terminals are set up in this way to explore the resource allocation in the ISAC system when the terminals have diverse demands.}, as shown in Fig. 1. 
\begin{figure}[h]
	\centerline{\includegraphics[width=8.5cm]{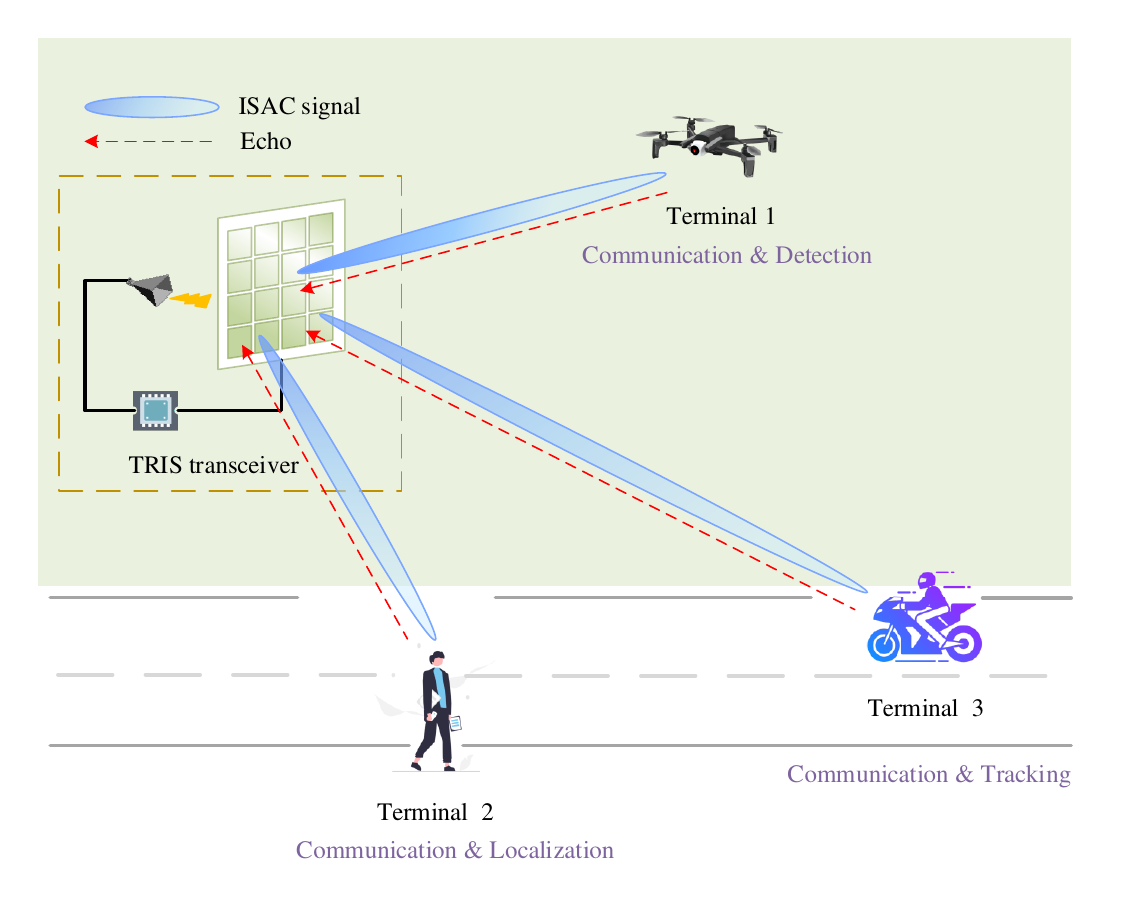}}
	\caption{TRIS transmitter empowered ISAC systems.}
\end{figure}
\subsection{TRIS Transmitter Characteristics}

The ISAC base station (BS) consists of horn antenna, TRIS and controller. The surface has $N = {N_r} \times {N_c}$ elements and forms a uniform planar array (UPA) with half wavelength spacing. In this structure, since the antenna and the terminals are located on different sides of the TRIS, the feed source obstruction problem and the self-interference of the reflected wave in the reflected RIS are solved. Furthermore, TRIS serves as a spatial diversity and loading information, which utilizes the control signals generated by the TMA to load precoded information onto the TRIS elements, and each element is loaded with different information, and the carrier wave penetrates the elements and carries the information to achieve direct modulation, thus realizing the function of traditional multiple antennas with lower energy consumption\cite{Liu2023}. Based on this feature, the communication and sensing streams are coded jointly, by combining them with TMA, which in turn facilitates the integration of sensing and multistream communication.

Modulating the message on the 1st harmonic will have an energy loss of 0.91 dB, and the power of each TRIS element loaded with the message cannot exceed the power of that harmonic\cite{he}, and the power constraints of the TRIS need to satisfy the following equation
\begin{equation}
		\setlength{\abovedisplayskip}{3pt}
	\setlength{\belowdisplayskip}{3pt}
{{{\left[ {{{\bf{P}}}{\bf{P}}^H} \right]}_{nn}}}  \le {P_t},\forall n,
\end{equation}
where ${\bf{P}}$ denotes the linear precoding matrix, and the maximum available power of each TRIS element is $P_t$.

\subsection{Channel Model} 
The channel can be categorized into a communication channel and a radar sensing channel according to whether the signal is traveling back and forth at the BS and the terminal. 

The communication channel ${{\bf{h}}_k} \in {\mathbb{C}^{N \times 1}}$ between the BS and the terminal $k$ is assumed to be known by the BS and terminal\footnote{There are numerous methods for obtaining channel state information, and in this paper we take the approach of channel estimation. In addition, sensing can also assist in obtaining channel state information, but in this paper more attention is paid to the tradeoff between sensing and communication.}, and the channel state information acquisition can be referred to in \cite{9732214,Wang2011}. Due to the mobility of the terminal, the fast fading and time-varying characteristics of the channel can be expressed in the following form
\begin{equation}
		\setlength{\abovedisplayskip}{3pt}
	\setlength{\belowdisplayskip}{3pt}
{{\bf{h}}_k} = {\rho  _k}{\bf{a}}\left( {{\theta _k},{\varphi _k}} \right)\delta \left( {t - {\tau _k}} \right){e^{j2\pi {f_k}lT}},\forall k,
\end{equation}
where ${\rho _k}$, ${\tau _k}$, ${f_k}$ and $T$ denote the path loss, one-way path delay, one-way Doppler shift, and symbol time, respectively, $l \in {\cal L} = \left\{ {1, \cdots ,L} \right\}$ denotes a discrete time index of a coherent processing interval (CPI), ${\bf{a}}\left( {{\theta _k},{\varphi _k}} \right)\in {\mathbb{C}^{N \times 1}}$ denotes the steering vector of the electromagnetic wave in the TRIS element w.r.t. the $k$-th target azimuth angle ${\varphi _k}$ and pitch angle ${\theta _k}$, which can be expressed as

\begin{equation}
		\setlength{\abovedisplayskip}{3pt}
	\setlength{\belowdisplayskip}{3pt}
{\bf{a}}\left( {{\theta _k},{\varphi _k}} \right) = {\left[ {{e^{ - j2\pi {\delta _r}{{\bf{n}}_r}}}} \right]^T} \otimes {\left[ {{e^{ - j2\pi {\delta _c}{{\bf{n}}_c}}}} \right]^T},\forall k,
\end{equation}
where ${{\bf{n}}_r} = \left[ {0,1, \cdots ,{N_r} - 1} \right]$, ${{\bf{n}}_c} = \left[ {0,1, \cdots ,{N_c} - 1} \right]$, ${\delta _r} = {{d_f}\sin {\theta _k}\cos {\varphi _k}}/{\lambda }$, and ${\delta _c} = {{d_f}\sin {\theta _k}\sin {\varphi _k}}/{\lambda }$. $\left( {{n_r},{n_c}} \right)$ denotes the element position index of the RIS, $\lambda $ denotes the wavelength, and ${d_f}$ denotes the center distance of adjacent RIS elements.

Considering a single BS radar, the signal is reflected back to the BS after reaching the target, and since there is only a single antenna for uplink reception, the radar sensing channel can be expressed as

\begin{equation}
		\setlength{\abovedisplayskip}{3pt}
	\setlength{\belowdisplayskip}{3pt}
{{\bf{g}}_k} = {\alpha _k}{\bf{a}}\left( {{\theta _k},{\varphi _k}} \right)b^H{\left( {{\theta _k},{\varphi _k}} \right)}\delta \left( {t - {\tau _{d,k}}} \right){e^{j2\pi {f_{d,k}}lT}},\forall k,
\end{equation}
where $b\left( {{\theta _k},{\varphi _k}} \right) = {e^{{{ - j2\pi d\left( {\sin {\theta _k}\cos {\varphi _k} + \sin {\theta _k}\sin {\varphi _k}} \right)} \mathord{\left/
				{\vphantom {{ - j2\pi d\left( {\sin {\theta _k}\cos {\varphi _k} + \sin {\theta _k}\sin {\varphi _k}} \right)} \lambda }} \right.
				\kern-\nulldelimiterspace} \lambda }}}$ denotes the steering scalar from the target to the receiving antenna. ${\alpha _k}$ denotes the complex reflection coefficient associated with the target radar cross-section, ${\alpha _k} \sim {\cal C}{\cal N}\left( {0,\sigma _{{\alpha _k}}^2} \right)$ and $\sigma _{{\alpha _k}}^2 = {S_{RCS}} {\lambda}^2 / {((4\pi)^3d^4)}$, where ${S_{RCS}}$ denotes the RCS of target, and ${\tau _{d,k}} = 2{\tau _k}$ and ${f_{d,k}} = 2{f_k}$ denote the round-trip time delay and Doppler frequency shift\cite{ding}, respectively.

\subsection{Signal Model} 

\begin{figure*}[htbp]
	\centering
	\includegraphics[scale=0.40]{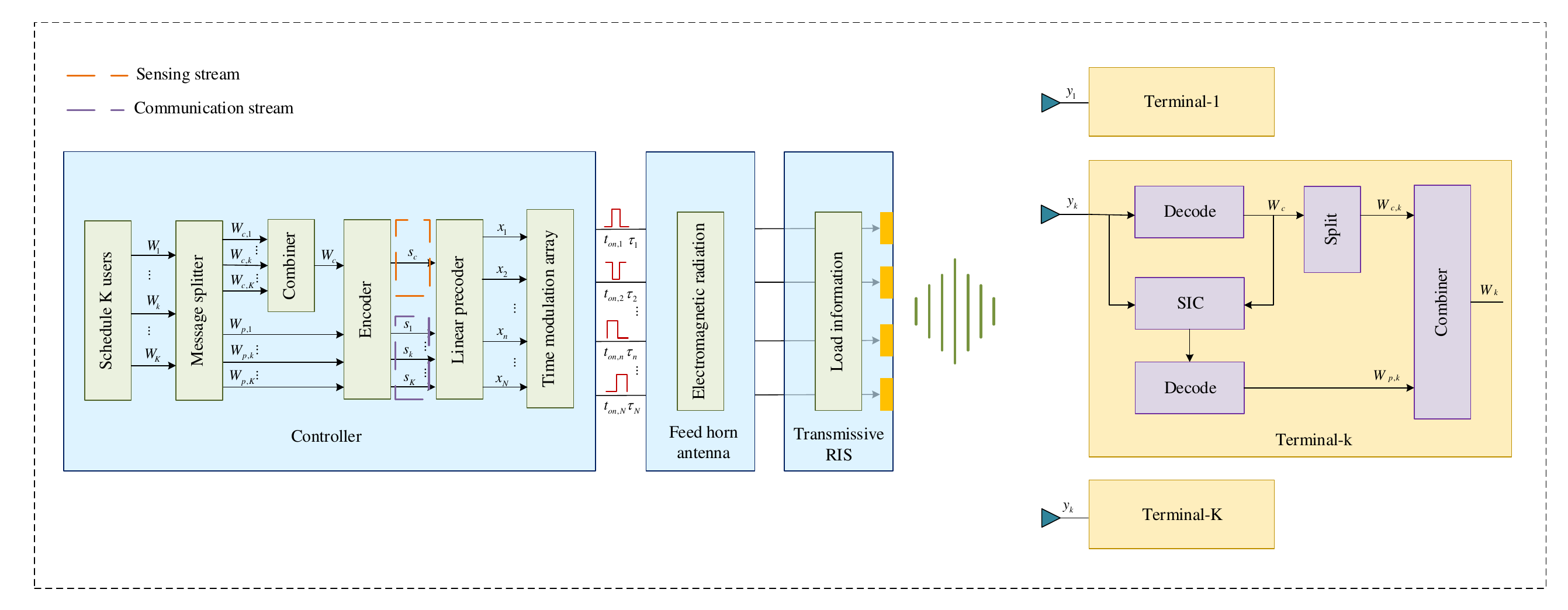}
	\caption{Encoding and decoding processes.}
\end{figure*}
In this paper, rate split signaling is considered where the transmission information is categorized into common and private streams. Both common and private streams are used for communication, in addition to utilizing common stream for sensing and designing it independently\footnote{In order to provide additional design freedom to the system while satisfying the communication requirements, we utilize the common stream for sensing, but in practice the private stream is also reflected back to the BS, so we constrain the reflected private stream power in equation (11).}. The signal can be expressed in the following form

\begin{equation}
		\setlength{\abovedisplayskip}{3pt}
	\setlength{\belowdisplayskip}{3pt}
	{\bf{x}}(t) = {\bf{P}}{{\bf{s}}(t)},
\end{equation}
where ${\bf{P}} = \left[ {{{\bf{p}}_c},{{\bf{p}}_1}, \cdots ,{{\bf{p}}_k}} \right] \in \mathbb{C}{^{N \times \left( {K + 1} \right)}}$ denotes the linear precoding matrix, and ${\bf{s}}(t) = {\left[ {{s_c(t)},{s_1(t)}, \cdots ,{s_K(t)}} \right]^T} \in \mathbb{C}{^{\left( {K + 1} \right) \times 1}}$ represents the transmission information, which is a wide stationary process in the time domain, statistically independent and with zero mean. In this paper, the common information of all users $ \left\{ {W_{c,1}, \cdots ,W_{c,K}} \right\}$ are combined as $W_c$ and jointly encoded into a common stream $s_c$ using the codebook codes shared by all users. The private information of all users $\left\{ {{W_{p,1}}, \cdots ,{W_{p,K}}} \right\}$ are encoded independently as private streams $\left\{{s_1},\cdots ,{s_K}\right\}$, which are only decoded by the corresponding terminal. The encoding and decoding processes are given by Fig. 2.

The signal received by the $k$-th terminal is expressed as

\begin{equation}
		\setlength{\abovedisplayskip}{3pt}
	\setlength{\belowdisplayskip}{3pt}
	{y_k\left(t\right)} = \sum\limits_{j \in {\cal \hat K}}{\bf{h}}_k^H{{{\bf{p}}_j}}{{s_j}\left(t-\tau _{k}\right)}+ {v_k\left(t\right)},\forall k,
\end{equation}
where ${v_k} \sim {\cal C}{\cal N}\left( {0,{\sigma _{{v_k}}^2}} \right)$ denotes the Gaussian white noise at the terminal and ${\cal \hat K}=\left\{c,1,...,K\right\}$.

The echo signal received by the BS can be expressed as
\begin{equation}
		\setlength{\abovedisplayskip}{3pt}
	\setlength{\belowdisplayskip}{3pt}
{r\left(t\right)} = \sum\limits_{k \in {K}}\left(\sum\limits_{j \in {\cal \hat K}}{\bf{g}}_k^H{{\bf{p}}_j}{s_j\left(t-\tau _{d,k}\right)} + {n_k\left(t\right)}\right),
\end{equation}
where ${n_k} \sim {\cal C}{\cal N}\left( {0,{\sigma _{{n_k}}^2}} \right)$ denotes the Gaussian white noise at the BS.
\section{Quality of Service Evaluation Metrics And Problem Formulation}
In this section, we will provide the communication and sensing QoSs. Based on this indicators, the corresponding optimization problem will be formulated in Section III-E.
\subsection{Communication Quality of Service}
In this paper, we consider achievable sum rate as performance criteria for terminal communication. When users decoding the sensing stream, the communication stream is considered as interference. After decoding the sensing stream, this part is removed in the received signal and the remaining terminal's communication stream is considered as interference for decoding the current terminal's information. Then, the corresponding sensing and communication stream signal-to-noise ratio of the $k$-th terminal can be expressed as follows

\begin{equation}
		\setlength{\abovedisplayskip}{3pt}
	\setlength{\belowdisplayskip}{3pt}
	{\gamma _{c,k}} = \frac{{{{\left| { {{\bf{h}}_k^H} {{\bf{p}}_c}} \right|}^2}}}{{\sum\nolimits_{i=1}^K {{{\left| {{{\bf{h}}_k^H} {{\bf{p}}_{i}}} \right|}^2}}  + {\sigma _{{v_k}}^2}}}, {\gamma _{p,k}} = \frac{{{{\left| {{\bf{h}}_k^H{{\bf{p}}_{k}}} \right|}^2}}}{{\sum\nolimits_{i \ne k}^K {{{\left| {{\bf{h}}_k^H{{\bf{p}}_{i}}} \right|}^2}}  + {\sigma _{{v_k}}^2}}},
\end{equation}
Then the corresponding achievable rate can be expressed as
\begin{equation}
	{R_{i,k}} = W{\log _2}\left( {1 + {\gamma _{i,k}}} \right), i \in \left\{{c,p}\right\}.
\end{equation}
The sensing stream is shared by all users and jointly participates in the encoding of the information. In order to ensure that all users are able to decode the sensing stream and satisfy the sensing service requirements, the following constraints need to be satisfied
\begin{equation}
	{R_c} = \min \left( {{R_{c,1}}, \cdots ,{R_{c,K}}} \right), ~{\rm{and}}~\sum\nolimits_{k=1}^K {{C_k}}  = {R_c},
\end{equation}
where ${C_k}$ denotes the equivalent common stream rate of the $k$-th terminal. As mentioned above, the private stream power reflected back to the BS needs to be constrained, which needs to satisfy
\begin{equation}
	\sum\limits_{i = 1}^K {{{\left| {{\bf{g}}_k^H{{\bf{p}}_i}} \right|}^2}}  \le {I_1}, \forall k,
\end{equation}
where ${I_1}$ denotes the interference threshold at the BS.

To satisfy the terminal's communication service requirements, we set the terminal sum rate threshold ${R_{{\rm{th}}}}$, and take the terminal's sum rate ${R}_{tot}$ greater than this threshold as the constraints of the optimization problem as follows
\begin{equation}
	{R}_{tot} = \sum\nolimits_{k=1}^K { \left({C_k+{R_{p,k}}} \right)}  \ge  R_{th}.
\end{equation}
\subsection{Target Detection}
Target detection is the detection of the presence of a target within a certain range of interest. Extensively, radar systems emit omni-directional signals to detect targets, which are not suitable for the network in this paper due to the complex environment resulting in a large amount of clutter interference and the requirement for detection services only in a small range of directions. Therefore, in this paper, we consider detecting the presence of terminal 1 in a specific area with known Doppler shift and time delay\footnote{The parameters are pre-calculated by the estimation algorithm\cite{Yin2022a,Sadi2012}. Regarding the acquisition of the Doppler shift, it can be obtained by utilizing the Fast Fourier Transform (FFT) on the echo signal\cite{Yin2022a}. For distance measure, we have $d_{k}= {c}\tau _{d,k}/{2}$. Therefore, the distance can be obtained by simply measuring the time delay\cite{Sadi2012}. }. Since each target has a different time delay and Doppler shift, the maximum value of the correlation function is obtained at the BS by correlation reception at the corresponding time delay and Doppler shift. 

Specifically, the BS takes measurements of the echo, with ${{\cal H}_0}$ indicating that 
the target is not present within the time delay and ${{\cal H}_1}$ indicating that the target is present within the time delay. For each of these measurements there are four decision conditions as shown in Table I. For each measurement there are two probabilities, the corresponding measurement signals and probability densities are shown in Table II.

\begin{table}[h]
	\caption{Decision Conditions}
	\begin{center}
\begin{tabular}{|c|c|c|}
	\hline
	\diagbox{Real Condition}{Decision} & ${{\cal H}_0}$ & ${{\cal H}_1}$  \\
	\hline
	${{\cal H}_0}$ & No Report & False Alarm   \\
	\hline
	${{\cal H}_1}$ & Miss Alarm & Detection \\   
	\hline
\end{tabular}
\end{center}
\end{table}

\begin{table}[h]
	\caption{Probability Densities}
	\begin{center}
	\begin{tabular}{|c|c|c|}
		\hline
		{Two Probabilities} & Measurement &	Probability Density  \\
		\hline
		${{\cal H}_0}$ & ${\hat x_1} = {w_1}$ & $P\left( {{{\hat x}_1}|{{\cal H}_0}} \right)$   \\
		\hline
		${{\cal H}_1}$ & ${\hat x_k} = {\alpha _1}{\bf{c}}\left( {{\theta _1},{\varphi _1}} \right){{\bf{p}}_c} + {w_1}$ & $P\left( {{{\hat x}_1}|{{\cal H}_1}} \right)$ \\   
		\hline
	\end{tabular}
\end{center}
\end{table}
{\noindent where ${\bf{c}}\left( {{\theta _1},{\varphi _1}} \right) = b\left( {{\theta _1},{\varphi _1}} \right){\bf{a}}^H{\left( {{\theta _1},{\varphi _1}} \right)}$, and ${w_1} \sim {\cal C}{\cal N}\left( {0,\sigma _r^2} \right)$ denotes processing noise.}

\begin{lemma} 
	{\rm  Let ${\hat x_1} = \int {r\left( t \right)s_c^*\left( {t - {\tau _{d,1}}} \right)} {e^{ - j2\pi {f_{d,1}}t}}dt$ denote the output of the receiver. The optimal detector under the Neyman-Pearson sense is likelihood ratio test (LRT), which is given by}
\setlength{\abovedisplayskip}{3pt}
\setlength{\belowdisplayskip}{3pt}
	\begin{equation}
		T=\left|\hat{x}_1\right|^2 \mathop \gtrless \limits_{{H}_0}^{{H}_1}  \delta,
	\end{equation}
	{\rm where $\delta$ is a threshold.}
\end{lemma}
 {\it Proof of \bf{Lemma 1}:} See Appendix A.$\hfill\blacksquare$

According to {\bf Lemma 1}, the statistics of the likelihood ratio test detector under the proposed architecture in this article obey the following distribution
\setlength{\abovedisplayskip}{3pt}
\begin{equation}
T \sim \left\{ {\begin{array}{*{20}{l}}
		{{\sigma _r^2}{\cal X}_{\left( 2 \right)}^2/{2},}&{{{\cal H}_0}}\\
		{\left( {\frac{{\sigma _r^2}}{2} + \frac{{\sigma _{{\alpha _k}}^2{{\left\| {{{\bf{p}}_c}} \right\|}^2}{{\left\| {{\bf{c}}\left( {{\theta _1},{\varphi _1}} \right)} \right\|}^2}}}{2}} \right){\cal X}_{\left( 2 \right)}^2,}&{{{\cal H}_1}}
\end{array}} \right.,\label{17}
\end{equation}
where ${\cal X}_{\left( 2 \right)}^2$ denotes the Chi-square distribution with degree of freedom 2. Based on Eq. (\ref{17}), the probability of false alarms can be expressed as
\begin{equation}
	\begin{split}
	{P_{FA}} = \Pr \left( {T > \delta |{{\cal H}_0}} \right) &= \Pr \left( {\frac{{\sigma _r^2}}{2}{\cal X}_{\left( 2 \right)}^2 > \delta } \right) \\&= \Pr \left( {{\cal X}_{\left( 2 \right)}^2 > \frac{{2\delta }}{{\sigma _r^2}}} \right),
	\end{split}
\end{equation}
The threshold $\delta $ can be given by $	\delta  = \frac{{\sigma _r^2}}{2}F_{{\cal X}_{\left( 2 \right)}^2}^{ - 1}\left( {1 - {P_{FA}}} \right)$, where $F_{{\cal X}_{\left( 2 \right)}^2}^{ - 1}$ denotes the inverse operation of the Chi-square distribution cumulative function.
Then the detection probability can be expressed as
	\setlength{\abovedisplayskip}{3pt}
\setlength{\belowdisplayskip}{3pt}
\begin{equation}
	{P_D} = \Pr \left( {T > \delta |{{\cal H}_1}} \right) = 1 - {F_{{\cal X}_{\left( 2 \right)}^2}}\left( {\frac{{{{2\delta } \mathord{\left/
						{\vphantom {{2\delta } {\sigma _r^2}}} \right.
						\kern-\nulldelimiterspace} {\sigma _r^2}}}}{{1 + {{\left\| {{{\bf{p}}_c}} \right\|}^2}{\varsigma _1}}}} \right),
\end{equation}
where ${F_{{\cal X}_{\left( 2 \right)}^2}}$  denotes the cumulative distribution function of the chi-square distribution. Here, the normalized sensing channel gain can be expressed as
\begin{equation}
{\varsigma _1} ={\sigma _{{\alpha _1}}^2{{\left\| {{\bf{c}}\left( {{\theta _1},{\varphi _1}} \right)} \right\|}^4}}/{{\sigma _r^2}}.
\end{equation}
Having defined the above detection probabilities, we define the QoS of detection as ${\beta _1} = {\left\| {{{\bf{p}}_c}} \right\|^2}{\varsigma _1}$. 
\subsection{Target Localization}
Radar target localization refers to the utilization of radar to measure the echo signals of a target and measure the target's parameters such as distance, velocity and angle. Distances and velocities have been discussed above and the angle information can be extracted from the received radar signal by utilizing the two-dimensional search in Capon estimate \cite{Xu2006} of $\alpha_k$. 

This paper focuses on the lower bound of the estimation error under the designed architecture and thus we consider the Cram\'{e}r-Rao bound (CRB) as an error metric for radar localization. The CRB gives a theoretical lower bound on the variance of the unbiased estimator, and the CRB matrix can be computed from ${\bf{CRB}} = {{\bf{I}}^{ - 1}}$, where $\bf I$ is the Fisher information matrix (FIM) w.r.t. the target estimation parameter ${{\bf{\xi }}_2} = {\left[ {{\tau _{d,2}},{f_{d,2}},{\theta _2},{\varphi _2}} \right]^T}$. Here ${\tau _{d,2}},{f_{d,2}},{\theta _2}$, and ${\varphi _2}$ denote the time delay, Doppler frequency shift, azimuth and pitch angle of terminal 2. ${{\bf{I}}_2}$ for the terminal 2 can be expressed in the following form\footnote{Prior to considering localization errors, the parameters of the terminal are estimated in advance, and since the common stream of RSMA is used for localization, the BS has known information about the common stream, and combined with the estimated time delay and Doppler shift parameters, the received signal $\hat r_2$ of the second terminal can be separated from $r(t)$, so that the formula in Eq. (19) can be used to calculate the FIM of the terminal 2 individually. When the parameters of multiple terminals need to be estimated jointly, the FIM can be calculated by referring to the scheme in literature \cite{4359542}.}
\begin{equation}
{{\bf{I}}_2} = \left[ {\begin{array}{*{20}{l}}
		{{I_{{\tau _{d,2}}{\tau _{d,2}}}}}&{{I_{{\tau _{d,2}}{f_{d,2}}}}}&{{I_{{\tau _{d,2}}{\theta _2}}}}&{{I_{{\tau _{d,2}}{\varphi _2}}}}\\
		{I_{{\tau _{d,2}}{f_{d,2}}}^H}&{{I_{{f_{d,2}}{f_{d,2}}}}}&{{I_{{f_{d,2}}{\theta _2}}}}&{{I_{{f_{d,2}}{\varphi _2}}}}\\
		{I_{{\tau _{d,2}}{\theta _2}}^H}&{I_{{f_{d,2}}{\theta _2}}^H}&{{I_{{\theta _2}{\theta _2}}}}&{{I_{{\theta _2}{\varphi _2}}}}\\
		{I_{{\tau _{d,2}}{\varphi _2}}^H}&{I_{{f_{d,2}}{\varphi _2}}^H}&{I_{{\theta _2}{\varphi _2}}^H}&{{I_{{\varphi _2}{\varphi _2}}}}
\end{array}} \right].
\end{equation}
As mentioned before, the reflected energy of the communication stream is considered as noise. Define ${L_2} = {\hat r_2} - {n_2}$ and we have

\begin{equation}
	{\left[ {{{\bf{I}}_2}} \right]_{i,j}} = \frac{2}{{\sigma _r^2}}{\mathop{\rm Re}\nolimits} \left\{ {\sum\limits_{l = 1}^L {\frac{{\partial {L^H_2}{{\left[ l \right]}}}}{{\partial {\xi _{2,i}}}}\frac{{\partial {L_2}\left[ l \right]}}{{\partial {\xi _{2,j}}}}} } \right\},{\rm{ }}i,j \in \left\{ {1,2,3,4} \right\}.
\end{equation}
The element-specific expression for ${{\bf{I}}_2}$ is as follows 
\begin{equation}
	{I_{{\tau _{d,2}}{\tau _{d,2}}}} = {4WA{{\left| {{\alpha _2}} \right|}^2}}\left( {\varepsilon \overline {{F^2}}  +8WT^3 {{\left( {\pi {f_{d,2}}} \right)}^2}} \right)/{{\sigma _r^2}},
\end{equation}
\begin{equation}
{I_{{\tau _{d,2}}{f_{d,2}}}} = {16{f_{d,2}}AT^2{{\left| {\pi{\alpha _2} W} \right|}^2}}/{{\sigma _r^2}},
\end{equation}
\begin{equation}
	{I_{{\tau _{d,2}}{\theta _2}}} = {8\pi {f_{d,2}}BW{{\left| {{\alpha _2}}T \right|}^2}}/{{\sigma _r^2}},
\end{equation}
\begin{equation}
	{I_{{\tau _{d,2}}{\varphi _2}}} = {8\pi {f_{d,2}}CW{{\left| {{\alpha _2}}T \right|}^2}}/{{\sigma _r^2}},
\end{equation}
\begin{equation}
	{I_{{f_{d,2}}{\theta _2}}} = {\pi BTL\left( {L + 1} \right){{\left| {{\alpha _2}} \right|}^2}}/{{\sigma _r^2}},
\end{equation}
\begin{equation}
	{I_{{f_{d,2}}{\varphi _2}}} = {\pi CTL\left( {L + 1} \right){{\left| {{\alpha _2}} \right|}^2}}/{{\sigma _r^2}},
\end{equation}
\begin{equation}
	{I_{{f_{d,2}}{f_{d,2}}}} ={2AL\left( {L + 1} \right)\left( {2L + 1} \right){{\left| {\pi{\alpha _2} T} \right|}^2}}/{({3\sigma _r^2})},
\end{equation}
\begin{equation}
	{I_{{\theta _2}{\theta _2}}} = {2{{\left| {{\alpha _2}} \right|}^2}DL}/{{\sigma _r^2}},
\end{equation}
\begin{equation}
	{I_{{\theta _2}{\varphi _2}}} ={2{{\left| {{\alpha _2}} \right|}^2}EL}/{{\sigma _r^2}},
\end{equation}
\begin{equation}
	{I_{{\varphi _2}{\varphi _2}}} = {2{{\left| {{\alpha _2}} \right|}^2}FL}/{{\sigma _r^2}}.
\end{equation}
For convenience, we have the following substitution
\begin{equation}
\left\{ {\begin{array}{*{20}{l}}
		{A = {\rm{Re}}\left\{ {{\bf{c}}\left( {{\theta _2},{\varphi _2}} \right){{\bf{R}}_{c,2}}{\bf{c}}^H{{\left( {{\theta _2},{\varphi _2}} \right)}}} \right\}}\\
		{B = {\rm{Re}}\left\{ { - j\frac{{\partial {\bf{c}}\left( {{\theta _2},{\varphi _2}} \right)}}{{\partial {\theta _2}}}{{\bf{R}}_{c,2}}{\bf{c}}^H{{\left( {{\theta _2},{\varphi _2}} \right)}}} \right\}}\\
		{C = {\rm{Re}}\left\{ { - j\frac{{\partial {\bf{c}}\left( {{\theta _2},{\varphi _2}} \right)}}{{\partial {\varphi _2}}}{{\bf{R}}_{c,2}}{\bf{c}}^H{{\left( {{\theta _2},{\varphi _2}} \right)}}} \right\}}\\
		{D = {\rm{Re}}\left\{ {\frac{{\partial {\bf{c}}\left( {{\theta _2},{\varphi _2}} \right)}}{{\partial {\theta _k}}}{{\bf{R}}_{c,2}}\frac{{\partial {\bf{c}}^H{{\left( {{\theta _2},{\varphi _2}} \right)}}}}{{\partial {\theta _2}}}} \right\}}\\
		{E = {\rm{Re}}\left\{ {\frac{{\partial {\bf{c}}\left( {{\theta _2},{\varphi _2}} \right)}}{{\partial {\theta _k}}}{{\bf{R}}_{c,2}}\frac{{\partial {\bf{c}}^H{{\left( {{\theta _2},{\varphi _2}} \right)}}}}{{\partial {\varphi _2}}}} \right\}}\\
		{F = {\rm{Re}}\left\{ {\frac{{\partial {\bf{c}}\left( {{\theta _2},{\varphi _2}} \right)}}{{\partial {\varphi _2}}}{{\bf{R}}_{c,2}}\frac{{\partial {\bf{c}}^H{{\left( {{\theta _2},{\varphi _2}} \right)}}}}{{\partial {\varphi _2}}}} \right\}}
\end{array}} \right.,
\end{equation}
where ${{\bf{R}}_{c,2}} = \frac{1}{L}\sum\limits_{l = 1}^L {{{\bf{x}}_c}\left[ {l - {\tau _{d,2}}} \right]} {{\bf{x}}^H_c}{\left[ {l - {\tau _{d,2}}} \right]} = {{\bf{p}}_c}{\bf{p}}_c^H{\bf R}_{\tau_{d,2}}$, ${\bf R}_{\tau_{d,2}}$ denotes the correlation matrix of the Doppler delay, and the partial derivatives can be expressed as

\begin{equation}
	\frac{{\partial {\bf{c}}\left( {{\theta _2},{\varphi _2}} \right)}}{{\partial {\theta _2}}} = \frac{{\partial b\left( {{\theta _2},{\varphi _2}} \right)}}{{\partial {\theta _2}}}{\bf{a}}^H{\left( {{\theta _2},{\varphi _2}} \right)} + b\left( {{\theta _2},{\varphi _2}} \right)\frac{{\partial {\bf{a}}^H{{\left( {{\theta _2},{\varphi _2}} \right)}}}}{{\partial {\theta _2}}},
\end{equation}
\begin{equation}
\frac{{\partial {\bf{c}}\left( {{\theta _2},{\varphi _2}} \right)}}{{\partial {\varphi _2}}} = \frac{{\partial b\left( {{\theta _2},{\varphi _2}} \right)}}{{\partial {\varphi _2}}}{\bf{a}}^H{\left( {{\theta _2},{\varphi _2}} \right)} + b\left( {{\theta _2},{\varphi _2}} \right)\frac{{\partial {\bf{a}}^H{{\left( {{\theta _2},{\varphi _2}} \right)}}}}{{\partial {\varphi _2}}}.
\end{equation}

See Appendix B for a detailed derivation.$\hfill\blacksquare$

Since minimizing the maximum eigenvalue of the CRB is equivalent to maximizing the minimum eigenvalue of the FIM, we take the minimum eigenvalue of the target's FIM as the QoS for target localization, i.e.,

\begin{equation}
	{\mu _2} = {\sigma _{\min }}\left( {{{\bf{I}}_2}} \right).\label{local}
\end{equation}
\subsection{Target Tracking}
In this section, we consider the problem of target tracking. Target tracking is the process of predicting the target's state at the next moment based on the estimated target state parameters. The Kalman filter is one of the most basic target tracking algorithms, which utilizes a state evolution model and a measurement model for predicting and updating the target state and evaluating the tracking accuracy through the state covariance matrix. Then, we construct a problem that utilizes the extended Kalman filter (EKF) to track the position, velocity, and angle of terminal 3.

{\noindent \bf State Evolution Model:}

We focus on the position and velocity tracking so that the other parameters can be computed, hence the state variable of the $m$-th tracking interval of the terminal 3 can be expressed as ${{\bf{e}}_{3,m}} = {\left[ {{x_{3,m}},{y_{3,m}},{z_{3,m}},{{\dot x}_{3,m}},{{\dot y}_{3,m}},{{\dot z}_{3,m}}} \right]^T}$, where $\left( {{x_{3,m}},{y_{3,m}},{z_{3,m}}} \right)$ and $\left( {{{\dot x}_{3,m}},{{\dot y}_{3,m}},{{\dot z}_{3,m}}} \right)$ denote the position and velocity of the terminal 3, respectively. Considering the constancy of the moving elements of the object, the state evolution model can be expressed in the following form
\begin{equation}
	{{\bf{e}}_{3,m}} = {\bf{F}}{{\bf{e}}_{3,m - 1}} + {{\bf{w}}_{3,m - 1}},
\end{equation}
where ${\bf{F}}$ and ${{\bf{w}}_{3,m - 1}}$ denote the transfer matrix and the processing noise, respectively, and ${{\bf{w}}_{3,m - 1}}$ obeys a zero-mean Gaussian distribution. The transfer matrix ${\bf{F}}$ and covariance matrix ${\Delta _3}$ can be expressed as
\begin{equation}
	{\bf{F}} = \left[ {\begin{array}{*{20}{c}}
			1&{{T_s}}\\
			0&1
	\end{array}} \right] \otimes {{\bf{E}}_3},
\end{equation}
and
\begin{equation}
	{\Delta _3} = {\tilde \sigma _3}\left[ {\begin{array}{*{20}{c}}
			{\frac{1}{3}T_s^3}&{\frac{1}{2}T_s^2}\\
			{\frac{1}{2}T_s^2}&{{T_s}}
	\end{array}} \right] \otimes {{\bf{E}}_3},
\end{equation}
where ${\tilde \sigma _3}$, ${T_s}$, and ${{\bf{E}}_3}$ denote the process noise, sampling interval and unit matrix, respectively.

{\noindent \bf Measurement Model:}

The nonlinear measurement model can be expressed as

\begin{equation}
{{\bf{q}}_{3,m}} = f\left( {{{\bf{e}}_{3,m}}} \right) + {{\bf{\tilde w}}_{3,m}},
\end{equation}
where ${{\bf{q}}_{3,m}} = {\left[ {{d_{3,m}},{v_{3,m}},{\theta _{3,m}},{\varphi _{3,m}}} \right]^T}$ denotes the estimated parameters of the radar and $f\left(  \cdot  \right)$ denotes the nonlinear processing transformation, which can be expressed in the following form

\begin{equation}
	f\left( {{{\bf{e}}_{3,m}}} \right) = \left\{ {\begin{array}{*{20}{l}}
			{{d_{3,m}} = \sqrt {x_{3,m}^2 + y_{3,m}^2 + z_{3,m}^2} }\\
			{{v_{3,m}} = {{\left( {{x_{3,m}}{{\dot x}_{3,m}} + {y_{3,m}}{{\dot y}_{3,m}} + {z_{3,m}}{{\dot z}_{3,m}}} \right)} \mathord{\left/
						{\vphantom {{\left( {{x_{3,m}}{{\dot x}_{3,m}} + {y_{3,m}}{{\dot y}_{3,m}} + {z_{3,m}}{{\dot z}_{3,m}}} \right)} {{d_{3,m}}}}} \right.
						\kern-\nulldelimiterspace} {{d_{3,m}}}}}\\
			{{\theta _{3,m}} = \arctan \left( {{{{y_{3,m}}} \mathord{\left/
							{\vphantom {{{y_{3,m}}} {{x_{3,m}}}}} \right.
							\kern-\nulldelimiterspace} {{x_{3,m}}}}} \right)}\\
			{{\varphi _{3,m}} = \arctan \left( {{{{z_{3,m}}} \mathord{\left/
							{\vphantom {{{z_{3,m}}} {\sqrt {x_{3,m}^2 + y_{3,m}^2} }}} \right.
							\kern-\nulldelimiterspace} {\sqrt {x_{3,m}^2 + y_{3,m}^2} }}} \right)}
	\end{array}} \right..
\end{equation}
The measurement noise ${{\bf{\tilde w}}_{3,m}}$ is modeled as zero-mean Gaussian white noise, and the covariance matrix ${{\bf{\Psi }}_{3,m}}$ can be expressed as follows

\begin{equation}
	{{\bf{\Psi }}_{3,m}} = {\rm{diag}}\left( {\sigma _{{d_{3,m}}}^2,\sigma _{{v_{3,m}}}^2,\sigma _{{\theta _{3,m}}}^2,\sigma _{{\varphi _{3,m}}}^2} \right),
\end{equation}
where $\sigma _{{d_{3,m}}}^2,\sigma _{{v_{3,m}}}^2,\sigma _{{\theta _{3,m}}}^2,$ and $\sigma _{{\varphi _{3,m}}}^2$ denote the CRB for range, velocity, azimuth, and pitch, respectively.

Given the measurements ${{\bf{q}}_{3,m}}$, based on Bayesian theory, the joint probability density distribution function of ${{\bf{e}}_{3,m}}$ and ${{\bf{q}}_{3,m}}$ can be expressed as

\begin{equation}
	p\left( {{{\bf{e}}_{3,m}},{{\bf{q}}_{3,m}}} \right) = p\left( {{{\bf{q}}_{3,m}}|{{\bf{e}}_{3,m}}} \right)p\left( {{{\bf{e}}_{3,m}}} \right),
\end{equation}
where $p\left( {{{\bf{q}}_{3,m}}|{{\bf{e}}_{3,m}}} \right)$ denotes the conditional probability density distribution function of ${{\bf{q}}_{3,m}}$ given ${{\bf{e}}_{3,m}}$. Let ${\bf{J}}\left( {{{\bf{e}}_{3,m}}} \right)$ denotes the posterior FIM, which can be expressed as

\begin{equation}
	\begin{split}
	{\bf{J}}\left( {{{\bf{e}}_{3,m}}} \right)=  - {{\rm{E}}_{{{\bf{e}}_{3,m}},{{\bf{q}}_{3,m}}}}\left[ {\frac{{{\partial ^2}\ln p\left( {{{\bf{e}}_{3,m}},{{\bf{q}}_{3,m}}} \right)}}{{\partial {\bf{e}}_{3,m}^2}}} \right],
	\end{split}
\end{equation}
Then, according to the joint probability density distribution function, the posterior FIM can be expressed as

\begin{equation}
	{\bf{J}}\left( {{{\bf{e}}_{3,m}}} \right) = {{\bf{J}}_P}\left( {{{\bf{e}}_{3,m}}} \right) + {{\bf{J}}_D}\left( {{{\bf{e}}_{3,m}}} \right),
\end{equation}
where ${{\bf{J}}_P}\left( {{{\bf{e}}_{3,m}}} \right)$ and ${{\bf{J}}_D}\left( {{{\bf{e}}_{3,m}}} \right)$ denote the a priori information FIM and data FIM, respectively. They can be calculated by

\begin{equation}
	\begin{split}
		{{\bf{J}}_P}\left( {{{\bf{e}}_{3,m}}} \right) &=  - {{\rm{E}}_{{{\bf{e}}_{3,m}}}}\left[ {\frac{{{\partial ^2}\ln p\left( {{{\bf{e}}_{3,m}}} \right)}}{{\partial {\bf{e}}_{3,m}^2}}} \right]\\
		&= {\left( {{\Delta _k} + {\bf{F}}{{\bf{J}}_P}{{\left( {{{\bf{e}}_{3,m - 1}}} \right)}^{ - 1}}{{\bf{F}}^T}} \right)^{ - 1}},
	\end{split}
\end{equation}
and
\begin{equation}
	\begin{split}
		{{\bf{J}}_D}\left( {{{\bf{e}}_{3,m}}} \right) &=  - {{\rm{E}}_{{{\bf{e}}_{3,m}},{{\bf{q}}_{3,m}}}}\left[ {\frac{{{\partial ^2}\ln p\left( {{{\bf{q}}_{3,m}}|{{\bf{e}}_{3,m}}} \right)}}{{\partial {{\bf{e}}_{3,m}}\partial {{\bf{q}}_{3,m}}}}} \right]\\
		&= {{\rm{E}}_{{{\bf{e}}_{3,m}}}}\left[ {{\bf{H}}_{3,m}^T{\bf{\Psi }}_{3,m}^{ - 1}{{\bf{H}}_{3,m}}} \right]\\
		&= {\left. {{\bf{H}}_{3,m}^T{\bf{\Psi }}_{3,m}^{ - 1}{{\bf{H}}_{3,m}}} \right|_{{{{\bf{e}}}_{3,m}} = {{{\bf{\hat e}}}_{\left. {3,m} \right|m - 1}}}}.
	\end{split}
\end{equation}
where ${{{\bf{\hat e}}}_{\left. {3,m} \right|m - 1}}$ denotes the prior state estimate for the $m$-th state given the $(m-1)$-th state and ${{\bf{H}}_{3,m}}$ denotes the Jacobi matrix w.r.t. the nonlinear transformation $f\left(  \cdot  \right)$. Then the initial FIM can be rewritten as

\begin{equation}
	\begin{split}
	{\bf{J}}\left( {{{\bf{e}}_{3,m}}} \right) &= {\left( {{\Delta _3} + {\bf{F}}{{\bf{J}}_P}{{\left( {{{\bf{e}}_{3,m - 1}}} \right)}^{ - 1}}{{\bf{F}}^T}} \right)^{ - 1}} \\  &~~~~~~~~~~+ {\left. {{\bf{H}}_{3,l}^T{\bf{\Psi }}_{3,m}^{ - 1}{{\bf{H}}_{3,m}}} \right|_{{{\bf{e}}_{3,m}} = {{{\bf{\hat e}}}_{\left. {3,m} \right|m - 1}}}}.
	\end{split}
\end{equation}
The specific derivation of the above FIM can be given by the literature \cite{Tichavsky1998,Glass2011}. Having obtained the above expression, the goal of the trace is to maximize the trace of the posteriori FIM, i.e.,
\begin{equation}
{\gamma _{3,m}} = {\rm{tr}}\left( {{\bf{J}}\left( {{{\bf{e}}_{3,m}}} \right)} \right).
\end{equation}
Notice that the above equation depends on the predicted state information, which is determined by the state evolution model and the target state. Therefore, to achieve better performance, we use the EKF scheme as {\bf Algorithm \ref{alg1}}.
\begin{algorithm}[h]
	\caption{The EKF Algorithm}
	\label{alg1}
	\begin{algorithmic}[1]
		\STATE {\bf{Initialization}}: ${{\bf{e}}_{3,0}}$, ${\bf{F}}$, ${{\bf{M}}_{3,0}}$, ${\Delta _3}$, ${\bf{\hat H}}_{3,0}$, ${{{\bf{\hat \Psi }}}_{3,0}}$, maximum number of iterations $M$ and iteration index $m = 0$.
		\REPEAT
		\STATE State Prediction: 
		
		${{\bf{\hat e}}_{\left. {3,m} \right|m - 1}} = {\bf{F}}{{\bf{e}}_{3,m - 1}}$.
		\STATE MSE Matrix Prediction: 
		
		${{\bf{\hat M}}_{\left. {3,m} \right|m - 1}} = {\bf{F}}{{\bf{M}}_{3,m - 1}}{{\bf{F}}^H} + {\Delta _3}.
		\label{mse}$
		\STATE Power Allocation: Noting that ${{\bf{M}}_{3,m - 1}}$ is equivalent to ${\bf{J}^{-1}}\left( {{{\bf{e}}_{3,m}}} \right)$, the power is allocated by solving the optimization problem subject to satisfying the tracking quality of service to obtain ${{\bf{M}}_{3,m - 1}}$.
		\STATE Kalman Gain Calculation: 
		
		${{\bf{K}}_{3,m}} = {{\bf{\hat M}}_{\left. {3,m} \right|m - 1}}{\bf{\hat H}}_{3,m}^H{\left( {{{{\bf{\hat \Psi }}}_{3,m}} + {{{\bf{\hat H}}}_{3,l}}{{{\bf{\hat M}}}_{\left. {3,m} \right|m - 1}}{\bf{\hat H}}_{3,m}^H} \right)^{ - 1}}.$
		\STATE State Tracking: 
		
		${{\bf{\hat e}}_{3,m}} = {{\bf{\hat e}}_{\left. {3,m} \right|m - 1}} + {{\bf{K}}_{3,m}}\left( {{{\bf{q}}_{3,m}} - f\left( {{{{\bf{\hat e}}}_{\left. {3,m} \right|m - 1}}} \right)} \right).$
		\STATE MSE Matrix Update: 
		
		${{\bf{M}}_{3,m}} = \left( {{\bf{E}} - {{\bf{K}}_{3,m}}{{{\bf{\hat H}}}_{3,m}}} \right){{\bf{\hat M}}_{\left. {3,m} \right|m - 1}}.$
		\STATE $m \leftarrow m + 1$.
		\UNTIL The maximum number of iterations is reached.
		\STATE {\bf{return}} ${{\bf{\hat e}}_{3,m}}$, ${{\bf{\hat e}}_{\left. {3,m} \right|m - 1}}$, ${{\bf{M}}_{3,m}}$, and ${{\bf{\hat M}}_{\left. {3,m} \right|m - 1}}$.
	\end{algorithmic}  
\end{algorithm}

In summary, the service quality evaluation metrics corresponding to the three terminals are given. In fact, the evaluation metrics for detection, localization and tracking can be easily extended to the multi-terminal scenario after replacing the index with $k$. Finally, in order to visualize in which direction the beam is strengthened, we calculate its beampattern as

\begin{equation}
	B\left( {{\theta _{j}},{\varphi _{j}}} \right) = {{\bf{a}}^H}\left( {{\theta _{j}},{\varphi _{j}}} \right){{\bf{p}}_c}{\bf{p}}_c^H{\bf{a}}\left( {{\theta _{j}},{\varphi _{j}}} \right),
\end{equation}
where $j$ denotes the $j$-th angle grid among all $J$ grids.
\subsection{Problem Formulation} 
Having defined the above QoSs, the following optimization problem is constructed using the Pareto model with the cooperation of sensing QoS as the objective function and the communication QoS as the constraint. Due to the dynamic characteristics of the target, a static optimization problem such as (P1) is first established. For the $m$-th round, we have
\begin{subequations}
	\begin{align}
		&\left( {{\rm{P1}}} \right){\rm{:~}}\mathop {{\rm{max}}}\limits_{{\bf{P}},{\bf{c}}} {\rm{~}} {{\beta _1}{\rm{ + }}{\mu _2}{\rm{ + }}{\gamma _{3,m}}}  \notag\\
		&~~~~~~~~~~{\rm{s}}{\rm{.t}}{\rm{.}}~~~{{\bf{p}}_c}\succeq 0,{\rm{ }}{{\bf{p}}_k}\succeq 0,\forall k,\label{C1}\\
		&~~~~~~~~~~~~~~~~{{{\left[ {{{\bf{P}}}{\bf{P}}^H} \right]}_{nn}}}  \le {P_t},\forall n,\label{C2}\\
		&~~~~~~~~~~~~~~~~R_{tot} \ge {R_{{\rm{th}}}},\label{C3}\\
		&~~~~~~~~~~~~~~~~{C_k} \ge 0,\forall k,\label{C4}\\
		&~~~~~~~~~~~~~~~~\sum\limits_{k = 1}^K {{C_k}}  \le {R_c},\label{C5}\\
		&~~~~~~~~~~~~~~~~\sum\limits_{i = 1}^K {{{\left| {{\bf{g}}_k^H{{\bf{p}}_i}} \right|}^2}}  \le {I_1},\forall k,\label{C7}
	\end{align}
\end{subequations}

It can be seen that the problem (P1) is a nonconvex problem, due to the coupling of variables and nonconvex operations. In the next section, we discuss how to transform, split and solve the problem.

\section{Joint Design of Communication and Sensing}
This section transforms the above non-convex problem into a convex one and solves it, while jointly designing the communication and sensing parameters, and finally gives the complexity and convergence analysis of the algorithm.
\subsection{Convex Transformation of the Formulated Problem} 
Let ${{\bf{F}}_k} = {{\bf{h}}_k}{\bf{h}}_k^H$, ${{\bf{G}}_k} = {{\bf{g}}_k}{\bf{g}}_k^H$, and ${{\bf{Q}}_{\hat k}} = {{\bf{p}}_{\hat k}}{\bf{p}}_{\hat k}^H$. $\forall\hat k \in \cal \hat K $ is the sensing stream and communication stream indexes.  $ {{\bf{Q}}_c}$ and ${\bf{Q}} = \left[{{{\bf{Q}}_1}, \cdots ,{{\bf{Q}}_K}} \right]$ denote the sensing stream matrix and the set of communication stream matrices, respectively.  Then the original problem (P1) can be reconstructed as
\begin{subequations}
\begin{align}
		&\left( {{\rm{P2}}} \right){\rm{:~}}\mathop {{\rm{max}}}\limits_{{{\bf{Q}}_c},{\bf{Q}},{\bf{c}},{{t}}} {\rm{ ~}}{ {{\beta _1}{\rm{ + }}{t}{\rm{ + }}{\gamma _{3,m}}} } \notag\\
	&~~~~~~~~~~~~~{\rm{s}}{\rm{.t}}{\rm{.}}~~~{\rm(\ref{C3})-(\ref{C5}),}\\
	&~~~~~~~~~~~~~~~~~~~~{\rm{rank}}\left( {{{\bf{Q}}_{\hat k}}} \right) = 1,\forall \hat k,\label{rank1}\\
	&~~~~~~~~~~~~~~~~~~~~{{\bf{Q}}_{\hat k}}\succeq0,\forall \hat k,\label{Q}\\
	&~~~~~~~~~~~~~~~~~~~~\sum\limits_{{\hat k} = 1}^{\hat K} {{\rm{tr}}\left( {{{\bf{Q}}_{\hat k}}} \right)}  \le {NP_t},\label{Power}\\
	&~~~~~~~~~~~~~~~~~~~~\sum\limits_{i = 1}^K {{\rm{tr}}\left( {{{\bf{G}}_k}{{\bf{Q}}_i}} \right)}  \le {I_1},\forall k,\label{I1}\\
	&~~~~~~~~~~~~~~~~~~~~{{\bf{I}}_2}\succeq{t}{\bf{E}_4}.\label{al}
\end{align}	
\end{subequations}
For the objective function, due to the high complexity of finding the minimum eigenvalue and the non-convex operation in ${\mu _2} = {\sigma _{\min }}\left( {{{\bf{I}}_2}} \right)$, we transform $\mu_2$ into its equivalent form, i.e., constraint (\ref{al}), and auxiliary variables ${t}$ are introduced, where ${\bf{E}_4}$ denotes the unitary matrix of dimension consistent with ${{\bf{I}}_2}$. It can be observed that the objective function in ${\gamma _{3,m}}$ is a convex function w.r.t. the optimization variable ${{\bf{Q}}_c}$, which we will analyze below. After obtaining the FIM ${{\bf{J}}_P}{\left( {{{\bf{e}}_{3,m - 1}}} \right)^{ - 1}}$ for $(m-1)$-th round, ${\left( {{\Delta _3} + {\bf{F}}{{\bf{J}}_P}{{\left( {{{\bf{e}}_{3,m - 1}}} \right)}^{ - 1}}{{\bf{F}}^T}} \right)^{ - 1}}$ is a constant term. Based on the expression for ${{\bf{\Psi }}_{3,l}}$, ${\left. {{\bf{H}}_{3,m}^T{\bf{\Psi }}_{3,m}^{ - 1}{{\bf{H}}_{3,m}}} \right|_{{{\bf{e}}_{3,l}} = {{{\bf{\hat e}}}_{\left. {3,m} \right|m - 1}}}}$ is a concave function w.r.t. ${{\bf{Q}}_c}$. Then ${\gamma _{3,m}}$ is a concave function.

For the processing of constraints, notice that constraint (\ref{C3}) is a non-convex constraint due to the presence of ${R_{p,k}}$ . It is transformed into the following form
	\setlength{\abovedisplayskip}{3pt}
\setlength{\belowdisplayskip}{3pt}
\begin{equation}
	\begin{split}
	{R_{p,k}}\left( {\bf{Q}} \right) &= W\log \left( {\sum\nolimits_{i = 1}^K {{\rm{tr}}\left( {{{\bf{F}}_k}{{\bf{Q}}_i}} \right)}  + {\sigma ^2}} \right)\\
	&~~~~~-W\log \left( {\sum\nolimits_{i \ne k}^K {{\rm{tr}}\left( {{{\bf{F}}_k}{{\bf{Q}}_i}} \right)}  + {\sigma ^2}} \right),\forall k.
	\end{split}
	\label{rpk}
\end{equation}
Since Eq. (\ref{rpk}) is still non-convex, utilizing the SCA, denote the first and second term as ${f_k}\left( {\bf{Q}} \right)$, ${v_k}\left( {\bf{Q}} \right)$, respectively. Perform the first order Taylor expansion of it as follows
\begin{equation}
	{v_k}{\left( {\bf{Q}} \right)^{ub}} \buildrel \Delta \over = {v_k}\left( {{{\bf{Q}}^{\left( r \right)}}} \right) + {\rm{vec}}{\left( {\nabla {v_k}\left( {{{\bf{Q}}^{\left( r \right)}}} \right)} \right)^T}{\rm{vec}}\left( {{\bf{Q}} - {{\bf{Q}}^{\left( r \right)}}} \right),
\end{equation}
where $\nabla {v_k}\left( {\bf{Q}} \right) = \left[ {\frac{{\partial {v_k}\left( {\bf{Q}} \right)}}{{\partial {{\bf{Q}}_1}}}, \cdots ,\frac{{\partial {v_k}\left( {\bf{Q}} \right)}}{{\partial {{\bf{Q}}_K}}}} \right]$ denotes the gradient of ${v_k}\left( {\bf{Q}} \right)$, ${\rm{vec}}$ denotes the matrix vectorization operation, and ${{{\bf{Q}}^{\left( r \right)}}}$ represents the value of ${\bf{Q}}$ for the $r$-th iteration. Then, the constraint (\ref{C3}) is transformed into convex constraint as

\begin{equation}
	\sum\limits_{k = 1}^K \left[C_k+{f_k}\left({\bf{Q}} \right) - {v_k}{\left( {\bf{Q}} \right)^{ub}}\right] \ge {R_{{\rm{th}}}},\label{COMQOS}
\end{equation}
It can be seen that constraint (\ref{C5}) is still not a non-convex set, and it can be transformed to

\begin{equation}
	{\rm{tr}}\left( {{{\bf{F}}_k}{{\bf{Q}}_c}} \right) \ge {\gamma _{c0}}\left( {\sum\limits_{i = 1}^K {{\rm{tr}}\left( {{{\bf{F}}_k}{{\bf{Q}}_i}} \right)}  + {\sigma ^2}} \right),\forall k,\label{co}
\end{equation}
where ${\gamma _{c0}} = {2^{{R_c}/W}} - 1$ denotes the signal-to-noise ratio corresponding to $\sum\limits_{k = 1}^K {{C_k}} $. 
Synthesizing the above statements, the problem (P2) can be reconstructed in the following form
\begin{subequations}
	\begin{align}
		&\left( {{\rm{P3}}} \right){\rm{:~}}\mathop {{\rm{max}}}\limits_{{{\bf{Q}}_c},{\bf{Q}},{\bf{c}},{{t}}} {\rm{ ~}} { {{\beta _1}{\rm{ + }}{t}{\rm{ + }}{\gamma _{3,m}}} } \notag\\
		&~~~~~~~~~~~~~{\rm{s}}{\rm{.t}}{\rm{.}}~~~{\rm(\ref{C4}),(\ref{rank1})-(\ref{al})},(\ref{COMQOS}),(\ref{co}).
	\end{align}	
\end{subequations}

It can be seen that problem (P3) is still a non-convex optimization problem due to the presence of the rank-one constraint (\ref{rank1}), which is treated in the next section for specific optimization problems.
\subsection{Problem Solving}
In this subsection, due to the coupling of variables, we split the problem (P3) into several subproblems to solve. The sensing stream precoding matrix ${\bf Q}_c$ is first optimized, given the common stream rate $\bf c$ and the set of communication stream precoding matrices $\bf Q$. The subproblem can be written as

\begin{subequations}
	\begin{align}
		&\left( {{\rm{P4}}} \right){\rm{:~}}\mathop {{\rm{max}}}\limits_{{{\bf{Q}}_c},{{t}}} {\rm{ ~}}{ {{\beta _1}{\rm{ + }}{t}{\rm{ + }}{\gamma _{3,m}}} } \notag\\
		&~~~~~~~~~~~~{\rm{s}}{\rm{.t}}{\rm{.}}~~~{\rm(\ref{Power}),(\ref{al}),(\ref{co})}\\
		&~~~~~~~~~~~~~~~~~~~{{\bf{Q}}_c} \succeq 0,\\
		&~~~~~~~~~~~~~~~~~~~{\rm rank}\left({{\bf Q}_c}\right)=1,\label{qc}
	\end{align}	
\end{subequations}
where constraint (\ref{qc}) is a rank-one constraint, which is a non-convex constraint and we use the SDR technique given in {\bf Remark 1} to relax it. The problem of being slack is a typical SDP and can be solved by applying the CVX toolbox \cite{cvxtool}.
\begin{remark}
	{\it From {\rm \cite{Huang2010}}, it is clear that the feasible solutions obtained after removing the rank-one constraints are all feasible for the problem before removing the constraints. The optimal solutions ${\bf{Q}}_c^ * $ obtained from the solution of the SDP problem are used to get ${\bf{p}}_c^ * $ by rank-one decomposition, if the rank of ${\bf{Q}}_c^ * $ is one. If it is not, Gaussian randomization is used to obtain an approximate solution.}
\end{remark}

Then the communication stream precoding matrix set $\bf Q$ is optimized based on the sensing stream precoding matrix ${\bf Q}_c$ obtained from problem (P4), which can be written in the following form
\begin{subequations}
	\begin{align}
		&\left( {{\rm{P5}}} \right){\rm{:~find~}}{\bf{Q}}\notag\\
		&~~~~~~~~~~{\rm{s}}{\rm{.t}}{\rm{.}}~{\rm (\ref{Power}),(\ref{I1}),(\ref{COMQOS}),(\ref{co})},\label{63b}\\
		&~~~~~~~~~~~~~~~{{\bf{Q}}_k} \succeq 0,\forall k,\label{63c}\\
		&~~~~~~~~~~~~~~~{\rm{rank}}\left( {{{\bf{Q}}_{k}}} \right) = 1,\forall k.
	\end{align}
\end{subequations}
Typically problem (P5) is solved by adopting {\bf Remark 1} for SDR relaxation of rank-one constraints. However, due to the need to solve the communication stream for $K$ users and the high number of TRIS elements, which results in a large problem solution size, it is difficult to return a rank-one solution through Gaussian randomization, so in this paper, the DC algorithm is used to transform the non-convex rank-one constraints and solve problem (P5).
\begin{prop}
{\rm For the positive semidefinite matrix ${\bf{M}} \in \mathbb{C}{^{N \times N}}$, ${\rm{tr}}\left( {\bf{M}} \right) > 0$, the rank-one constraint can be expressed as the difference between two convex functions, i.e.,}
\begin{equation}
	{\rm{rank}}\left( {\bf{M}} \right) = 1 \Leftrightarrow {\rm{tr}}\left( {\bf{M}} \right) - {\left\| {\bf{M}} \right\|_2} = 0,
\end{equation}
{\rm where ${\rm{tr}}\left( {\bf{M}} \right) = \sum\limits_{n = 1}^N {{\sigma _n}\left( {\bf{M}} \right)}$, ${\left\| {\bf{M}} \right\|_2} = {\sigma _1}\left( {\bf{M}} \right)$ is spectral norm, and ${\sigma _n}\left( {\bf{M}} \right)$ represents the $n$-th largest singular value of matrix $\bf M$.}	
\end{prop}
According to {\bf Proposition 1}, problem (P5) can be transformed into Problem (P5.1), which can be expressed as

\begin{subequations}
	\begin{align}
		&\left( {{\rm{P5.1}}} \right)~\mathop {{\rm{min}}}\limits_{{{\bf{Q}}_k}} ~{\rm{tr}}\left( {{{\bf{Q}}_k}} \right) - {\left\| {{{\bf{Q}}_k}} \right\|_2}\notag\\
		&~~~~~~~~~~~{\rm{s}}{\rm{.t}}{\rm{.}}~~{\rm (\ref{63b}),(\ref{63c})}.\label{65b}
	\end{align}
\end{subequations}
Since $- {\left\| {{{\bf{Q}}_k}} \right\|_2}$ is a concave function w.r.t. ${{\bf{Q}}_k}$, problem (P5.1) is still a non-convex optimization problem, where SCA is used to perform Taylor first-order expansion of $- {\left\| {{{\bf{Q}}_k}} \right\|_2}$ to obtain its upper bound expression. This can be written as equation (\ref{66}) as shown at the top of the following page, where ${{\bf{u}}_{\max }}\left( {{\bf{Q}}_k^{\left( r \right)}} \right)$ denotes the eigenvector corresponding to the largest singular value of the matrix ${{\bf{Q}}_k}$ at the $r$-th iteration. Thus, problem (P5.1) can be approximated by transforming it into problem (P5.2), which can be expressed as

\setcounter{equation}{59} 
\begin{subequations}
	\begin{align}
		&\left( {{\rm{P5.2}}} \right)~\mathop {{\rm{min}}}\limits_{{{\bf{Q}}_k}} ~{\rm{tr}}\left( {{{\bf{Q}}_k}} \right) + \left\| {{{\bf{Q}}_k}} \right\|_2^{{\rm{ub}}}\notag\\
		&~~~~~~~~~~~{\rm{s}}{\rm{.t}}{\rm{.}}~~{\rm (\ref{65b})}.
	\end{align}
\end{subequations}
The problem is a typical SDP problem that can be solved using the CVX toolbox \cite{cvxtool}. Due to the presence of $K$ terminals, it is necessary to continuously solve the problem (P5.2) $K$ times to ultimately obtain the communication stream precoding matrix.

Next, we optimize the common stream rate $\bf c$ based on the sensing stream precoding matrix $\bf Q_c$ and the communication stream precoding matrix $\bf Q$ obtained from problems (P4) and (P5.2), which is a feasible check problem that can be written in the following form

\begin{subequations}
\begin{align}
	&\left( {{\rm{P6}}} \right){\rm{:~find~}}{\bf{c}}\notag\\
	&~~~~~~~~~~{\rm{s}}{\rm{.t}}{\rm{.}}~(\ref{C3})-(\ref{C5}).
	\end{align}
\end{subequations}
Note that this is a linear programming problem that can be easily solved with the CVX toolbox \cite{cvxtool}.

As above mentioned, (P4), (P5.2) and (P6) are alternately iterated to find the optimal solution to problem (P3) for the $l$-epoch, which can be summarized as {\bf Algorithm \ref{alg2}} as follows. Finally, synthesizing {\bf Algorithm \ref{alg1}} and {\bf Algorithm \ref{alg2}}, the communication and sensing joint design algorithm proposed in this paper is shown in {\bf Algorithm \ref{alg3}}.

\begin{algorithm}[htbp]
	\caption{Joint Communication and Sensing Optimization Algorithm}
	\label{alg2}
	\begin{algorithmic}[1]
		\STATE {\bf{Initialization}}: ${\bf{Q}}_c^{\left( 0 \right)}$, ${{\bf{Q}}^{\left( 0 \right)}}$, ${{\bf{c}}^{\left( 0 \right)}}$, ${{{t}}^{\left( 0 \right)}}$, convergence threshold $\varepsilon $ and iteration index $r = 0$.
		\REPEAT
		\STATE Obtain sensing stream precoding matrix ${\bf{Q}}_c^{\left( r \right)}$, and auxiliary variable ${{{t}}^{\left( r \right)}}$ by solving problem (P4).
		\STATE Obtain communication stream precoding matrix set ${{\bf{Q}}^{\left( r \right)}}$ by solving problem (P5.2) $k$ times.
		\STATE Obtain common stream rate vector ${{\bf{c}}^{\left( r \right)}}$ by solving problem (P6).
		\STATE $r \leftarrow r + 1$.
		\UNTIL The fractional decrease of the objective value is
		below a threshold ${\varepsilon}$.
		\STATE {\bf{return}} Precoding matrix ${\bf{P}}$ and sensing stream vector $\bf c$.
	\end{algorithmic}  
\end{algorithm}

\begin{algorithm}[htbp]
	\caption{Overall Joint Communication and Sensing Design Algorithm}
	\label{alg3}
	\begin{algorithmic}[1]
		\STATE {\bf{Initialization}}: ${{\bf{e}}_{3,0}}$, ${\bf{F}}$, ${{\bf{M}}_{3,0}}$, ${\Delta _3}$, ${\bf{\hat H}}_{3,0}$, ${{{\bf{\hat \Psi }}}_{3,0}}$, ${\bf{Q}}_c^{\left( 0 \right)}$, ${{\bf{Q}}^{\left( 0 \right)}}$, ${{\bf{c}}^{\left( 0 \right)}}$, ${{{t}}^{\left( 0 \right)}}$, internal convergence threshold $\varepsilon $, internal iteration index $r = 0$, external maximum number of iterations $M$, and external iteration index $m = 0$.
		\REPEAT
		\STATE Obtain sensing stream precoding matrix ${\bf{Q}}_c^{\left( r_{end},m \right)}$, auxiliary variable ${{\bf{t}}^{\left( r_{end},m \right)}}$, and communication stream precoding matrix set ${{\bf{Q}}^{\left( r_{end},m \right)}}$ according to the {\bf Algorithm \ref{alg2}} based on the state information.
		\STATE Obtain posterior state estimation ${{\bf{\hat e}}_{3,m}}$, prior state estimation ${{\bf{\hat e}}_{\left. {3,m} \right|m - 1}}$, MSE ${{\bf{M}}_{3,m}}$, and prior MSE ${{\bf{\hat M}}_{\left. {3,m} \right|m - 1}}$ according to the {\bf Algorithm \ref{alg1}} based on the ${\bf{J}^{-1}}\left( {{{\bf{e}}_{3,m}}} \right)$ obtained in {\bf Algorithm \ref{alg2}}. 
		\STATE Update the state information.
		\STATE $m \leftarrow m + 1$.
		\UNTIL The maximum number of external iterations $M$ reaches.
		\STATE {\bf{return}} Precoding matrix ${\bf{P}}$, sensing stream vector $\bf c$, posterior state estimation ${{\bf{\hat e}}_{3,m}}$, prior state estimation ${{\bf{\hat e}}_{\left. {3,m} \right|m - 1}}$, mean square error ${{\bf{M}}_{3,m}}$, and prior mean square error ${{\bf{\hat M}}_{\left. {3,m} \right|m - 1}}$.
	\end{algorithmic}  
\end{algorithm}


\subsection{Convergence and Computational Complexity Analysis}
{\it 1) Computational Complexity Analysis:} For {\bf Algorithm \ref{alg2}}, problem (P4) is solved with complexity ${\cal O}\left( {K{N^{3.5}}} \right)$, problem (P5.2) is solved with complexity ${\cal O}\left( {K{N^{3.5}}} \right)$, and problem (P6) is solved with complexity ${\cal O}\left( K \right)$ in each iteration. Then, the computational complexity of {\bf Algorithm \ref{alg2}} is ${\cal O}\left( {K\log \left( {1/\varepsilon } \right)\left( {{2N^{3.5}} + 1} \right)} \right)$, where $\varepsilon $ denotes the precision of stopping the iteration, which is set to ${10^{ - 3}}$. For {\bf Algorithm \ref{alg1}}, the complexity is ${\cal O}\left( ({6KM})^{2} \right)$ in each iteration. Therefore, the overall computational complexity for {\bf Algorithm \ref{alg3}} is ${\cal O}\left( M({K\log \left( {1/\varepsilon } \right)\left( {{2N^{3.5}} + 1} \right)}+({6KM})^{2}) \right)$ \cite{CVX}.

{\it 2) Convergence Analysis:} The convergence analysis of joint communication and sensing optimization can be proved as follows. Firstly, analyze the convergence of {\bf Algorithm \ref{alg2}}. We define ${\bf{Q}}_c^{\left( r \right)}$, ${{\bf{Q}}^{\left( r \right)}}$, ${{{t}}^{\left( r \right)}}$, and ${{\bf{c}}^{\left( r \right)}}$ as the solutions of the $r$-th iteration of problems (P4), (P5.2) and (P6), so the objective function of the $r$-th iteration can be expressed as ${\cal F}\left( {{\bf{Q}}_c^{\left( r \right)},{{\bf{Q}}^{\left( r \right)}},{{{t}}^{\left( r \right)}},{{\bf{c}}^{\left( r \right)}}} \right)$.
In step 3 of {\bf Algorithm \ref{alg2}}, the precoding matrix and the auxiliary variables can be obtained given the communication stream precoding matrix set ${{\bf{Q}}^{\left( r \right)}}$ and common stream rate vector ${{\bf{c}}^{\left( r \right)}}$. Then there are
	\setlength{\abovedisplayskip}{3pt}
\setlength{\belowdisplayskip}{3pt}
\begin{equation}
	\setlength{\abovedisplayskip}{3pt}
	\setlength{\belowdisplayskip}{3pt}
		\begin{split}
		&{\cal F}\left( {{\bf{Q}}_c^{\left( r \right)},{{\bf{Q}}^{\left( r \right)}},{{{t}}^{\left( r \right)}},{{\bf{c}}^{\left( r \right)}}} \right) \\
		&~~~~~~~~~~~~~~~~~~~~\le {\cal F}\left( {{\bf{Q}}_c^{\left( {r + 1} \right)},{{\bf{Q}}^{\left( r \right)}},{{{t}}^{\left( {r + 1} \right)}},{{\bf{c}}^{\left( r \right)}}} \right).
		\end{split}
\end{equation}
In step 4 of {\bf Algorithm \ref{alg2}}, the communication stream precoding matrix can be obtained with the sensing stream precoding matrix ${\bf{Q}}_c^{\left( {r + 1} \right)}$ and auxiliary variables ${{{t}}^{\left( {r + 1} \right)}}$ given, then there are
	\setlength{\abovedisplayskip}{3pt}
\setlength{\belowdisplayskip}{3pt}
\begin{equation}
	\setlength{\abovedisplayskip}{3pt}
	\setlength{\belowdisplayskip}{3pt}
	\begin{split}
	&{\cal F}\left( {{\bf{Q}}_c^{\left( {r + 1} \right)},{{\bf{Q}}^{\left( r \right)}},{{{t}}^{\left( {r + 1} \right)}},{{\bf{c}}^{\left( r \right)}}} \right) \\
	&~~~~~~~~~~~~~~~~~~~\le {\cal F}\left( {{\bf{Q}}_c^{\left( {r + 1} \right)},{{\bf{Q}}^{\left( {r + 1} \right)}},{{{t}}^{\left( {r + 1} \right)}},{{\bf{c}}^{\left( r \right)}}} \right).
	\end{split}
\end{equation}
Similarly, in step 5 of {\bf Algorithm \ref{alg2}}, the common stream rate vector can be obtained with the sensing stream precoding matrix ${\bf{Q}}_c^{\left( {r + 1} \right)}$, the set of communication sensing precoding matrix ${{\bf{Q}}^{\left( {r + 1} \right)}}$, and the auxiliary variables ${{{t}}^{\left( {r + 1} \right)}}$ given, then there are
\newcounter{TempEqCnt}                         
\setcounter{TempEqCnt}{\value{equation}} 
\setcounter{equation}{58}                           
\begin{figure*} 
	\centering
	\begin{equation}
		- {\left\| {{{\bf{Q}}_k}} \right\|_2} \le  - \left( {{{\left\| {{\bf{Q}}_k^{\left( r \right)}} \right\|}_2} + {\rm{tr}}\left( {{{\bf{u}}_{\max }}\left( {{\bf{Q}}_k^{\left( r \right)}} \right){{\bf{u}}_{\max }}{{\left( {{\bf{Q}}_k^{\left( r \right)}} \right)}^H}\left( {{{\bf{Q}}_k} - {\bf{Q}}_k^{\left( r \right)}} \right)} \right)} \right) \buildrel \Delta \over = {\left( {{{\left\| {{{\bf{Q}}_k}} \right\|}_2}} \right)^{ub}},	\label{66}
	\end{equation}
		\hrulefill
\end{figure*}
\setcounter{equation}{\value{TempEqCnt}} 
\setlength{\abovedisplayskip}{3pt}
\setlength{\belowdisplayskip}{3pt}
\begin{equation}
	\setlength{\abovedisplayskip}{3pt}
	\setlength{\belowdisplayskip}{3pt}
	\begin{split}
	&{\cal F}\left( {{\bf{Q}}_c^{\left( {r + 1} \right)},{{\bf{Q}}^{\left( {r + 1} \right)}},{{{t}}^{\left( {r + 1} \right)}},{{\bf{c}}^{\left( r \right)}}} \right) \\
	&~~~~~~~~~~~~~~~~~\le {\cal F}\left( {{\bf{Q}}_c^{\left( {r + 1} \right)},{{\bf{Q}}^{\left( {r + 1} \right)}},{{{t}}^{\left( {r + 1} \right)}},{{\bf{c}}^{\left( {r + 1} \right)}}} \right).
	\end{split}
\end{equation}
Based on the above, there is ultimately
	\setlength{\abovedisplayskip}{3pt}
\setlength{\belowdisplayskip}{3pt}
\begin{equation}
	\setlength{\abovedisplayskip}{3pt}
	\setlength{\belowdisplayskip}{3pt}
	\begin{split}
	&{\cal F}\left( {{\bf{Q}}_c^{\left( r \right)},{{\bf{Q}}^{\left( r \right)}},{{{t}}^{\left( r \right)}},{{\bf{c}}^{\left( r \right)}}} \right) \\
	&~~~~~~~~~~~~~~~~~\le {\cal F}\left( {{\bf{Q}}_c^{\left( {r + 1} \right)},{{\bf{Q}}^{\left( {r + 1} \right)}},{{{t}}^{\left( {r + 1} \right)}},{{\bf{c}}^{\left( {r + 1} \right)}}} \right).
 	\end{split}
\end{equation}
This shows that at each iteration of {\bf Algorithm \ref{alg2}}, the objective function is non-decreasing. Since the objective function must be finite-valued upper bound, convergence of {\bf Algorithm \ref{alg2}} is guaranteed. The convergence of {\bf Algorithm \ref{alg1}} and {\bf Algorithm \ref{alg3}} is guaranteed due to the setting of the maximum number of iterations. Therefore, the proposed algorithm is convergent.
\section{Numerical Results}
In this section, numerical simulations of the proposed algorithm are performed to demonstrate its effectiveness. A three-dimensional Cartesian coordinate system is used, the BS is located at (0m, 0m, 0m), the terminals are distributed in the three-dimensional space with the BS as the coordinate origin at the start coordinates (50m, 55m, 50m), (-70m, -50m, 25m), and (50m, 100m, 50m) and they have velocities (5m/s,5m/s,0m/s), (0m/s,10m/s,0m/s) and (-10m/s,0m/s,0m/s) respectively. The remaining parameters are given in Table III.
\begin{table}[!htbp]
	\caption{Simulation Parameters}
	\begin{center}
\begin{tabular}{|l|c|}  
	\hline
	\bf{Parameters} & \bf{Value}\\ 
	\hline
	Carrier frequency ($f$) & 30 GHz \\
	\hline
	Maximum transmissive power ($P_{t}$) & 1 mW \\
	\hline
	System bandwidth ($W$) & 20 MHz \\
	\hline
	Minimum QoS of communication (${R_{th}}$) & 1 Mbps \\
	\hline
	Communication stream interference threshold (${I_1}$) & -60 dBm \\
	\hline
	Noise power (${\sigma ^2}$) & -90 dBm \\
	\hline
\end{tabular}
	\end{center}
\end{table}

\begin{figure}[!htbp]
	\centering
	\includegraphics[scale=0.45]{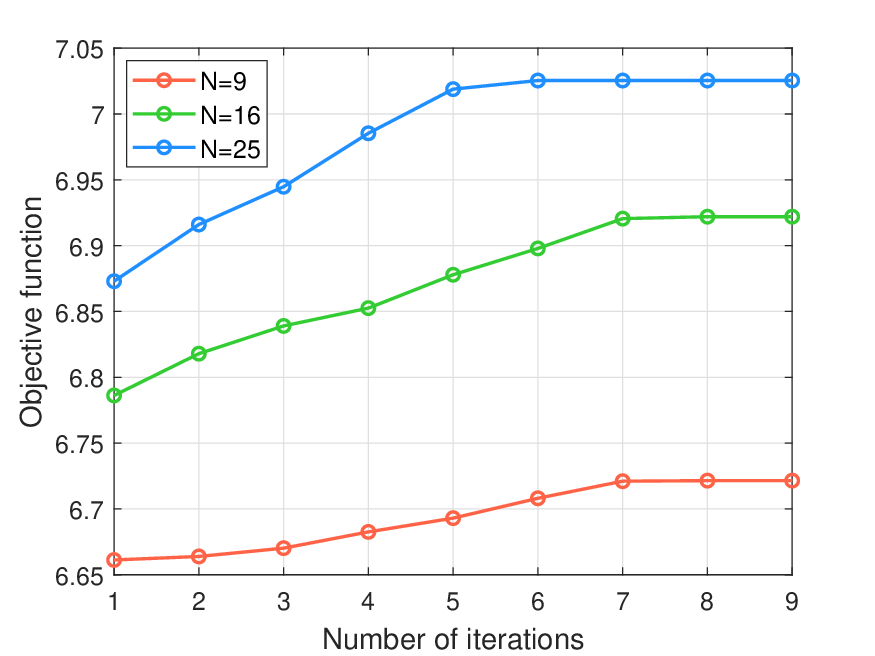}
	\caption{The convergence of the proposed joint communication and sensing optimization algorithm $({R_{th}}=1Mbits)$.}
	\label{con}
\end{figure}

Firstly, we verify the convergence of the proposed joint communication and sensing optimization algorithm. Fig. \ref{con} depicts the objective function value varies with the number of iterations under different TRIS elements. It is obvious that the algorithm can achieve better convergence performance in about 6 iterations. Moreover, we compare the effects of different TRIS elements on system performance. Specifically, comparing the value of the objective function with the number of TRIS elements of 9, 16 and 25, it can be seen that the larger the number of TRIS elements, the larger the value of the objective function, which indicates that the sensing ability of the system increases with the number of TRIS elements.

\begin{figure}[!htbp]
	\centering
	\begin{minipage}[t]{1.0\linewidth}
		\centering
		\begin{tabular}{@{\extracolsep{\fill}}c@{}c@{}@{\extracolsep{\fill}}}
			\includegraphics[width=0.525\linewidth]{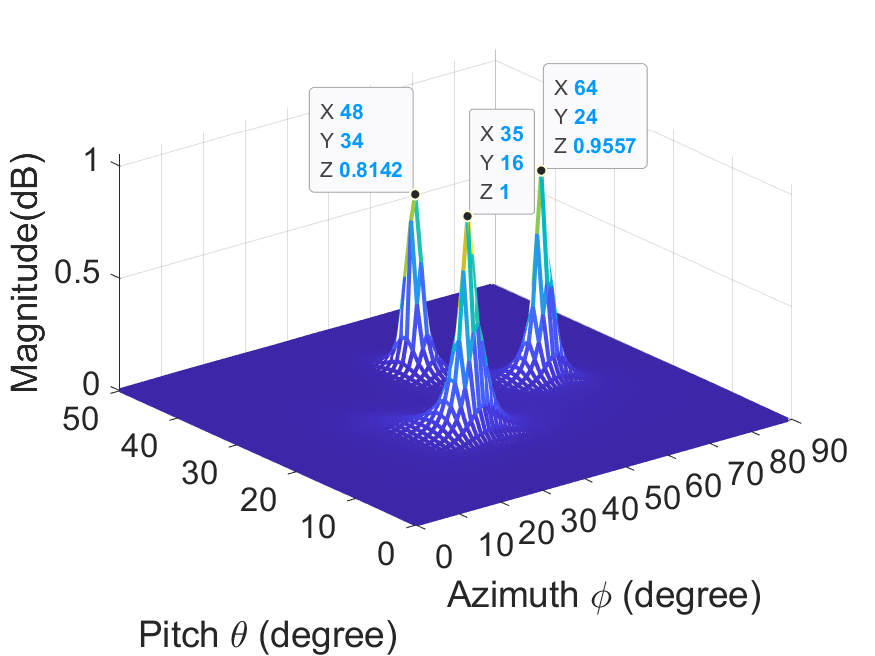}&
			\includegraphics[width=0.525\linewidth]{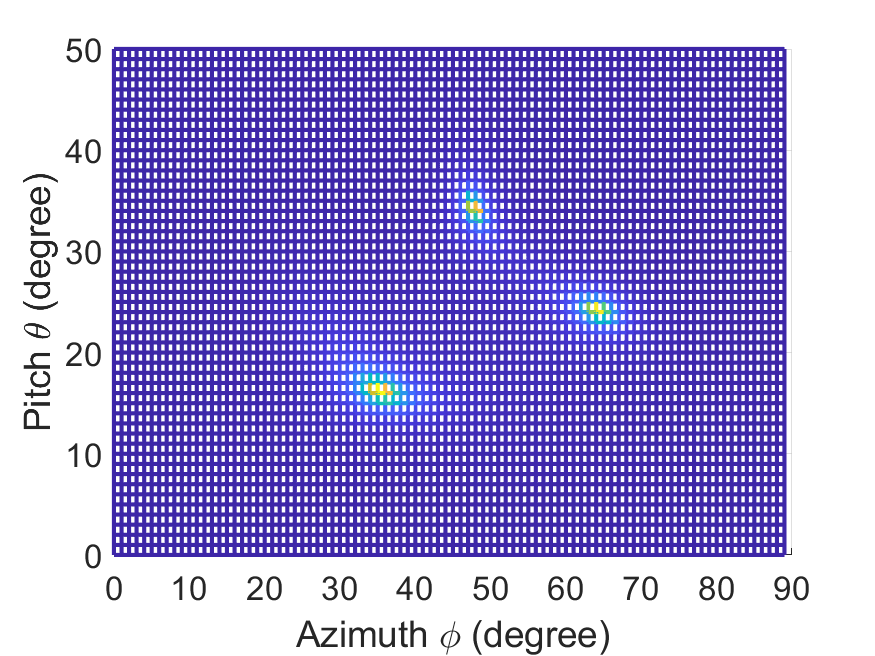}\\
			(a) 3D view & (b) Vertical view\\
		\end{tabular}
	\end{minipage}
	\begin{minipage}[t]{1.0\linewidth}
		\centering
		\begin{tabular}{@{\extracolsep{\fill}}c@{}c@{}@{\extracolsep{\fill}}}
			\includegraphics[width=0.525\linewidth]{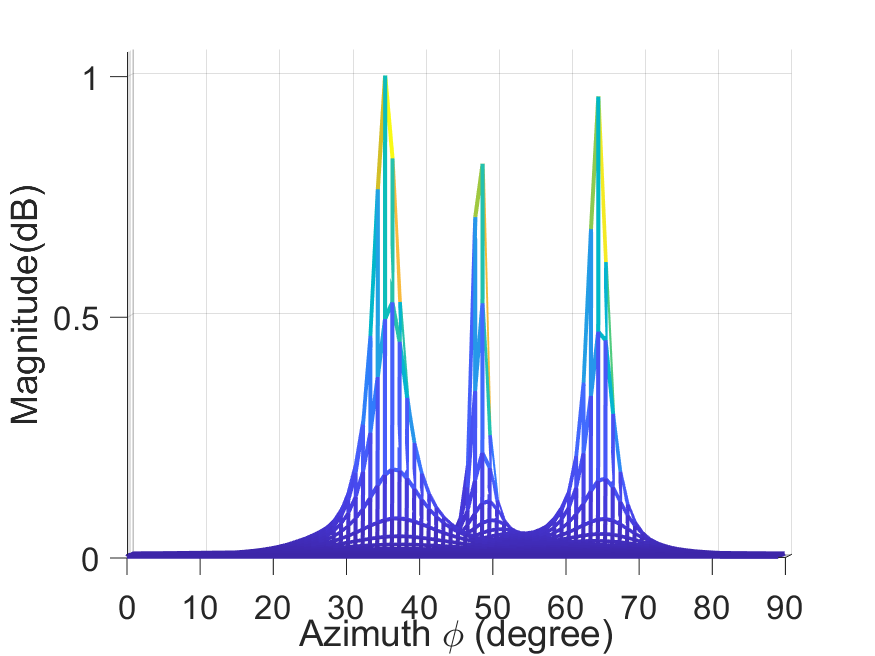} &
			\includegraphics[width=0.525\linewidth]{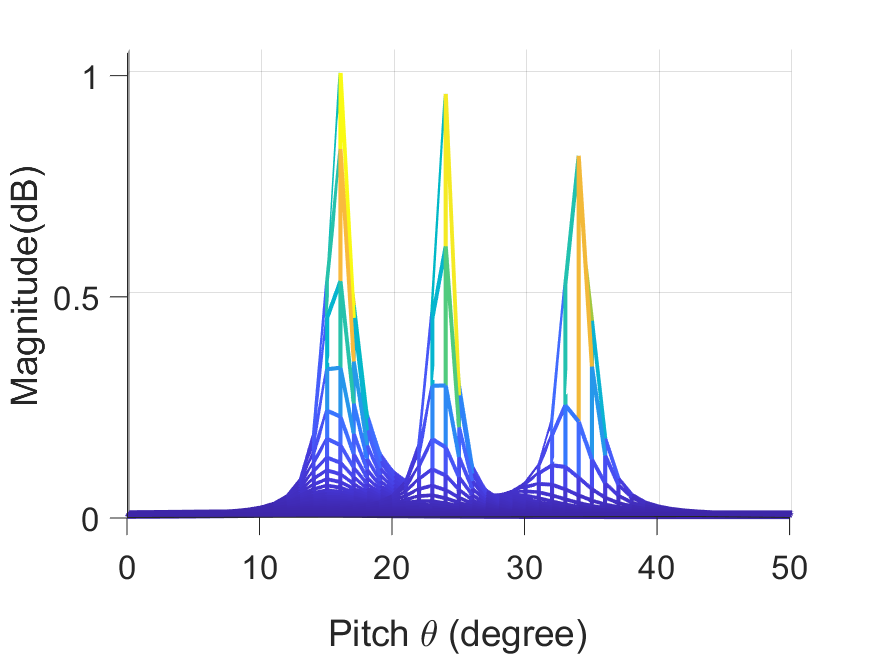}\\
			(c) Azimuth & (d) Pitch\\
		\end{tabular}
	\end{minipage}
	\begin{minipage}[t]{1.0\linewidth}
		\centering
		\begin{tabular}{@{\extracolsep{\fill}}c@{}@{\extracolsep{\fill}}}
			\includegraphics[width=1\linewidth]{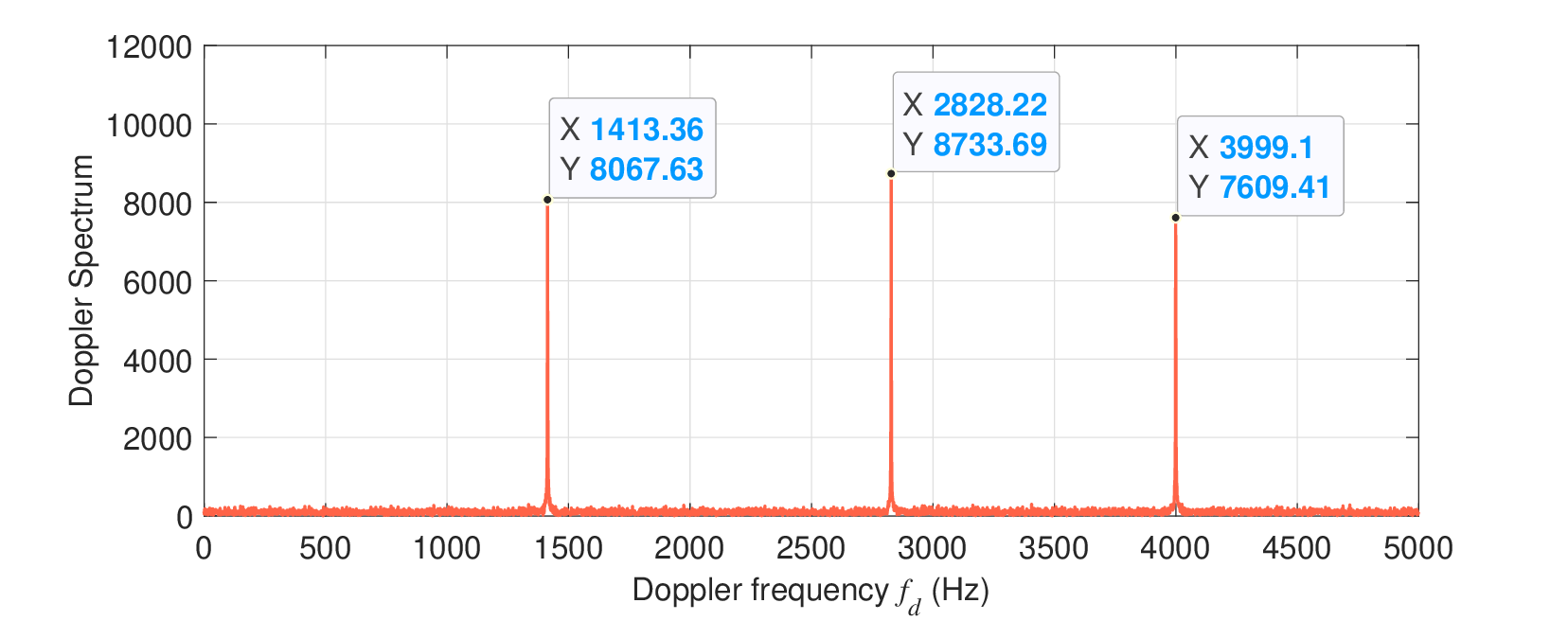}\\
			(e) Doppler frenquency\\
		\end{tabular}
	\end{minipage}
	\caption{An example of angle and Doppler frequency estimation. ($N=16, R_{th}=1Mbits, L=1024, f_s=2f_c$)}
	\label{Estimation}
\end{figure}

Secondly, a case study on angle and Doppler shift estimation is presented. By utilizing the two-dimensional search in Capon estimate, we can obtain $\theta$ and $\phi$.  Figs. \ref{Estimation} (a)-(d) present the corresponding angle estimations. By applying FFT to the echo signal, the Doppler spectrum of the terminals can be obtained as shown in Fig. \ref{Estimation} (e), and the Doppler shifts of the them are 1413.38 Hz, 2828.26 Hz and 3999.17 Hz, respectively.
\begin{figure}[!htbp]
	\centering
	\includegraphics[scale=0.45]{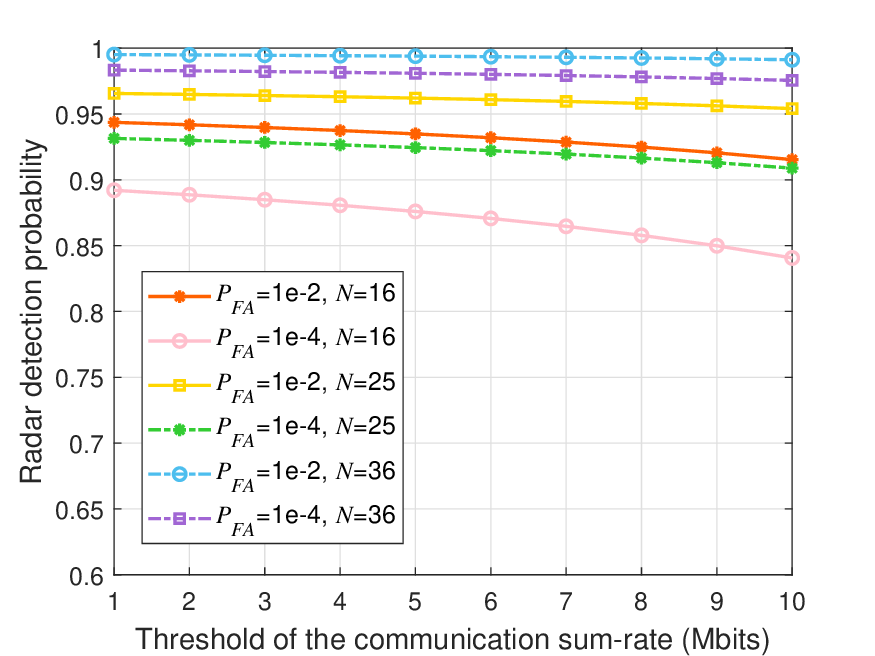}
	\caption{Detection Performance: The detection probability $P_D$ versus communication threshold under different TRIS elements and $P_{FA}$}
	\label{de}
\end{figure}

Thirdly, we illustrate the detection probability varies with the sum-rate threshold under different false alarm probabilities. Fig. \ref{de} shows that the detection probability decreases as the sum-rate threshold increases, which indicates the trade-off between the detection and communication. The detection probability is overwhelmingly above 80$\%$ for communication sum-rate of 10 Mbits, demonstrating the effectiveness of the architecture in ISAC network. In addition, as the false alarm probability increases, the probability of detection increases. Meanwhile, TRIS plays a key role in target detection, and Fig. \ref{de} shows that with the increase of TRIS elements, the detection probability is greatly improved under different false alarm probabilities.

\begin{figure}[!htbp]
	\centering
	\begin{minipage}[t]{1.0\linewidth}
		\centering
		\begin{tabular}{@{\extracolsep{\fill}}c@{}c@{}@{\extracolsep{\fill}}}
			\includegraphics[width=0.525\linewidth]{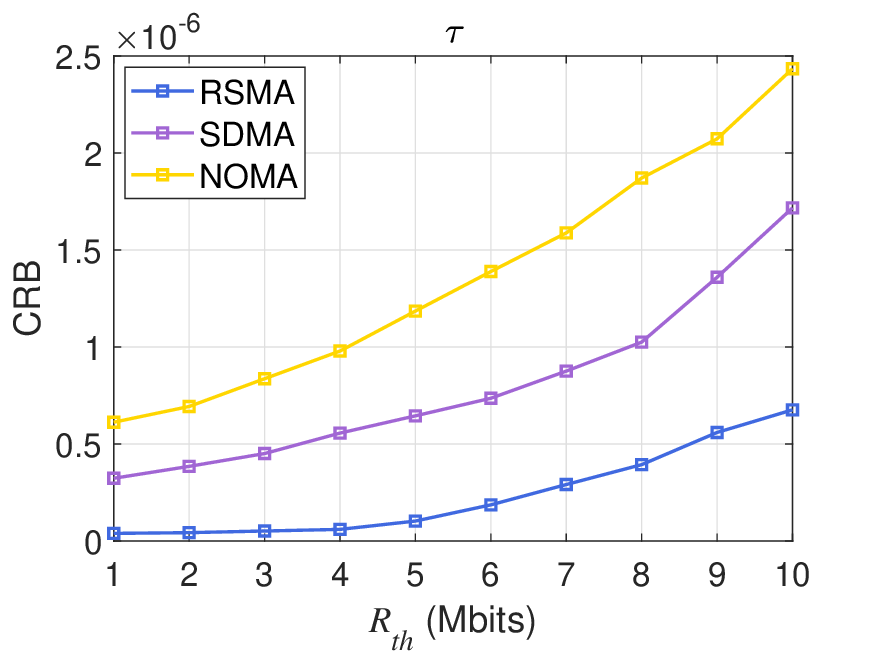}&
			\includegraphics[width=0.525\linewidth]{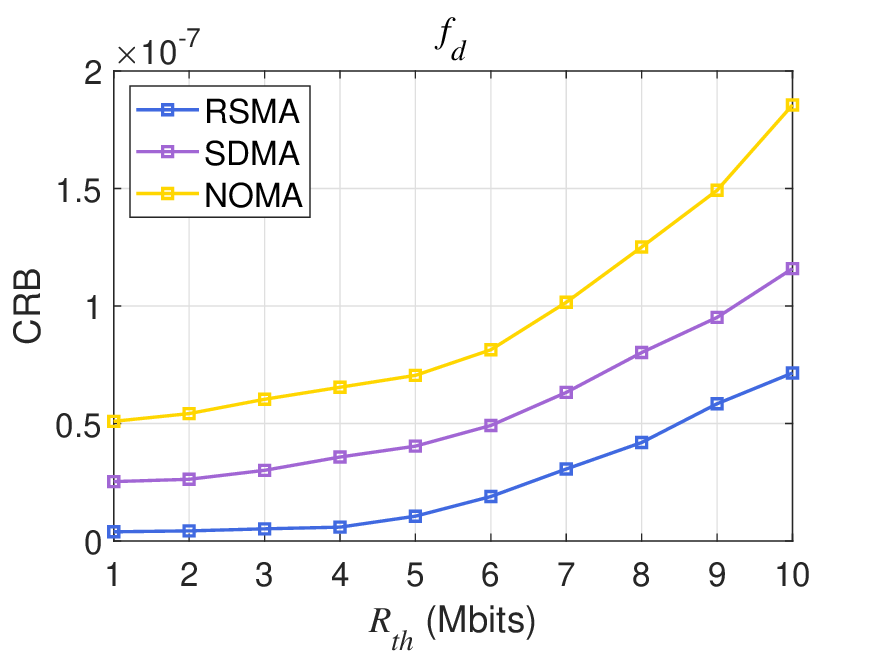}\\
			(a) Time delay & (b) Doppler frequency shift\\
		\end{tabular}
	\end{minipage}
	\begin{minipage}[t]{1.0\linewidth}
		\centering
		\begin{tabular}{@{\extracolsep{\fill}}c@{}c@{}@{\extracolsep{\fill}}}
			\includegraphics[width=0.525\linewidth]{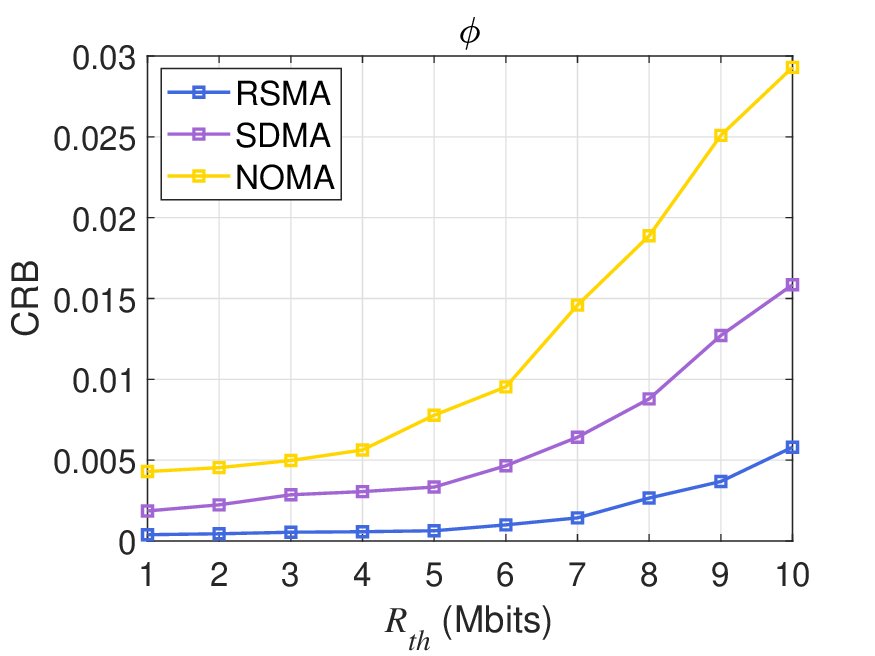} &
			\includegraphics[width=0.525\linewidth]{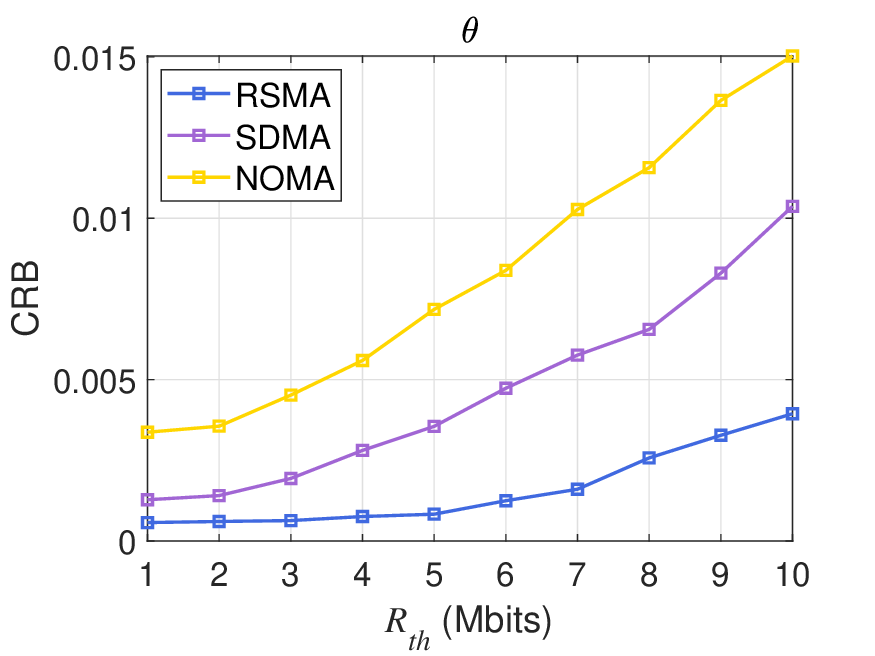}\\
			(c) Azimuth & (d) Pitch\\
		\end{tabular}
	\end{minipage}
	\caption{Localization Performance: CRB versus $R_{th}$ ($N=16$).}
	\label{CRB}
\end{figure}

\begin{figure*} [htbp]
	\centering
	\subfloat[\label{fig:a}]{
		\includegraphics[scale=0.40]{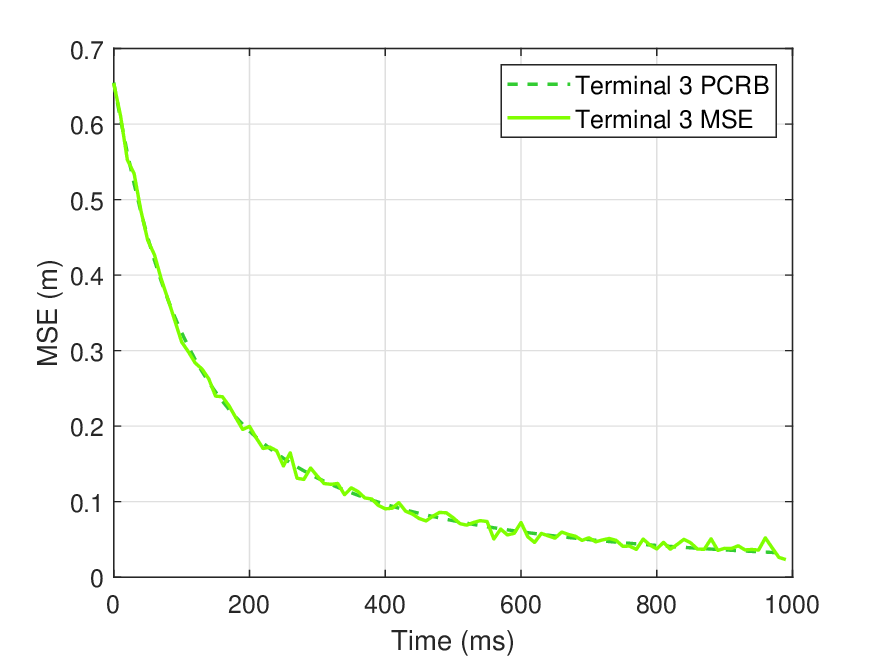}}
	\subfloat[\label{fig:b}]{
		\includegraphics[scale=0.40]{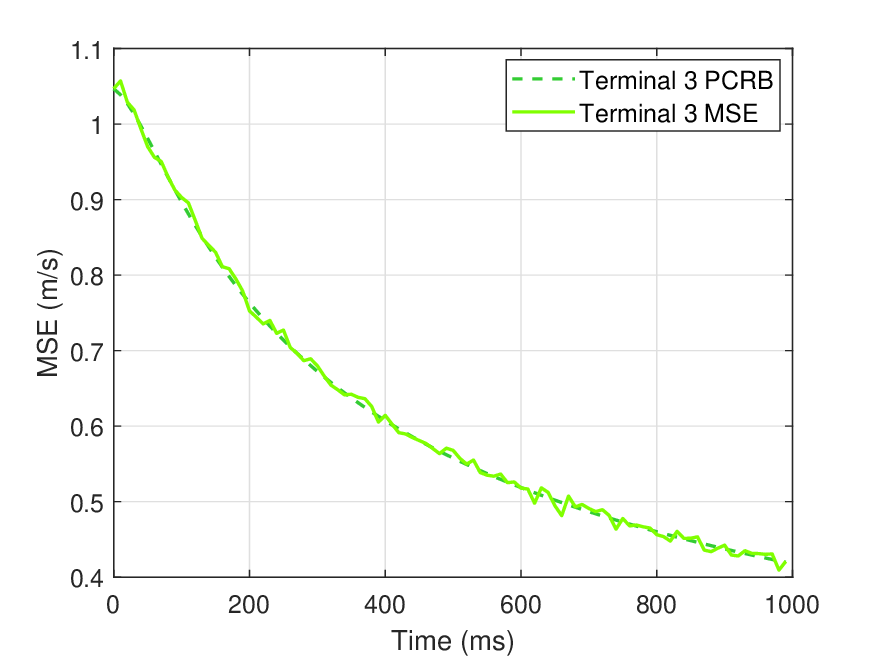}}
	\subfloat[\label{fig:c}]{
		\includegraphics[scale=0.40]{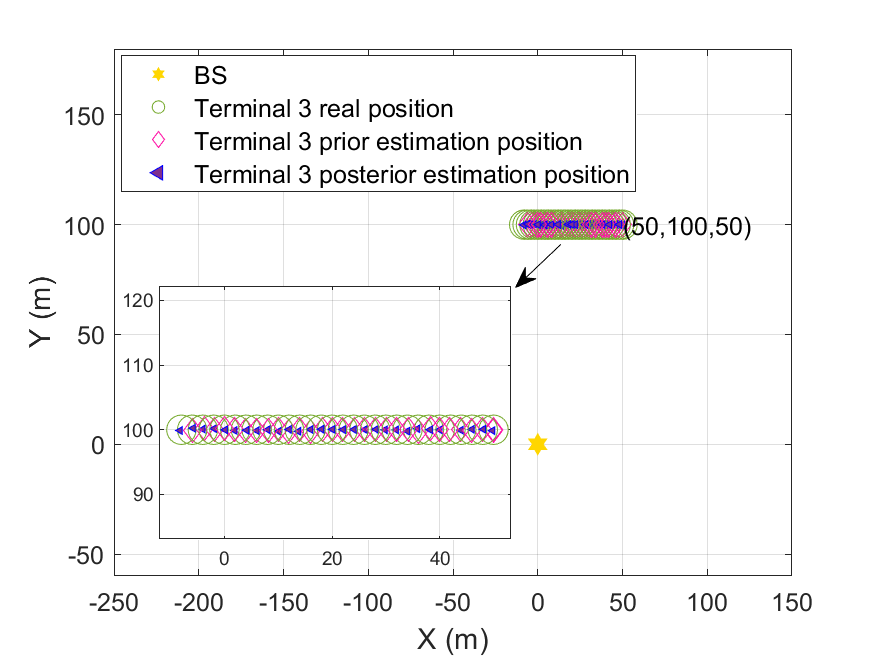}}
	\caption{Tracking Performance: (a) Distance tracking performance $({R_{th}}=1Mbits, N=16)$. (b) Velocity tracking performance $({R_{th}}=1Mbits, N=16)$. (c) Tracking trajectory map $({R_{th}}=1Mbits, N=16)$.}
	\label{tra}
\end{figure*}

Next, we study the target localization performance under this system. The CRB is analyzed as the sum-rate threshold changes. The benchmarks are selected as SDMA and NOMA, where SDMA can be considered as a special case of RSMA with the common stream turned off, and NOMA involves the decoding order. Fig. \ref{CRB} illustrates the CRB varies with the sum-rate threshold. The results demonstrate that RSMA outperforms SDMA and NOMA in terms of the CRB of time delay, doppler frequency shift, azimuth, and pitch, which indicates RSMA has a better ability to manage interference and adapts radar sensing than SDMA and NOMA. This is mainly due to the fact that under the design of this paper, the common and private streams of RSMA are designed independently and used for different functions, with more flexible interference management and without complex decoding order. From the curve trend, it can be seen that there is a trade-off between the improvement of communication service quality and the sensing error. Thus,  the specific design can reasonably choose the priority according to the scenario and service requirements. 

Then, we present the performance of target tracking. We utilize $T_s=0.02s$ as the block duration and the total tracking period is $2s$.  Fig. \ref{tra} (a) and (b) illustrate the MSE and PCRB for distance and velocity, and the results show that the MSE and PCRB decrease during this period. At the same time, PCRB and MSE largely overlap, proving the effectiveness of the algorithm for error prediction. Meanwhile, the trajectory of the terminal 3 can be given by Fig. \ref{tra} (c), where the priori estimated position, and the posteriori estimated position basically coincide with the real position, proving the reliability of the tracking algorithm.

\begin{figure}[H]
	\centering
	\begin{minipage}[t]{1.0\linewidth}
		\centering
		\begin{tabular}{@{\extracolsep{\fill}}c@{}c@{}@{\extracolsep{\fill}}}
			\includegraphics[width=0.525\linewidth]{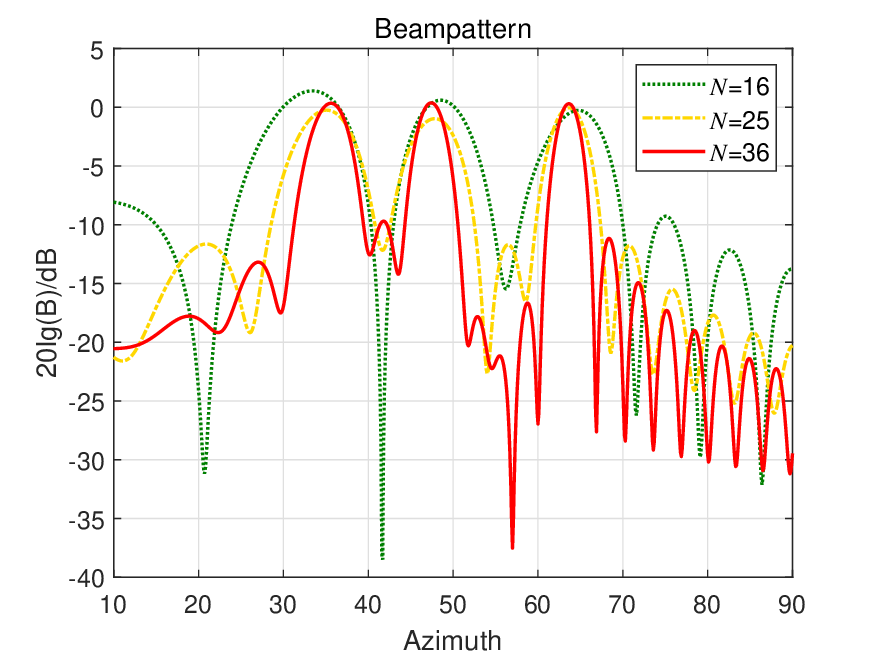}&
			\includegraphics[width=0.525\linewidth]{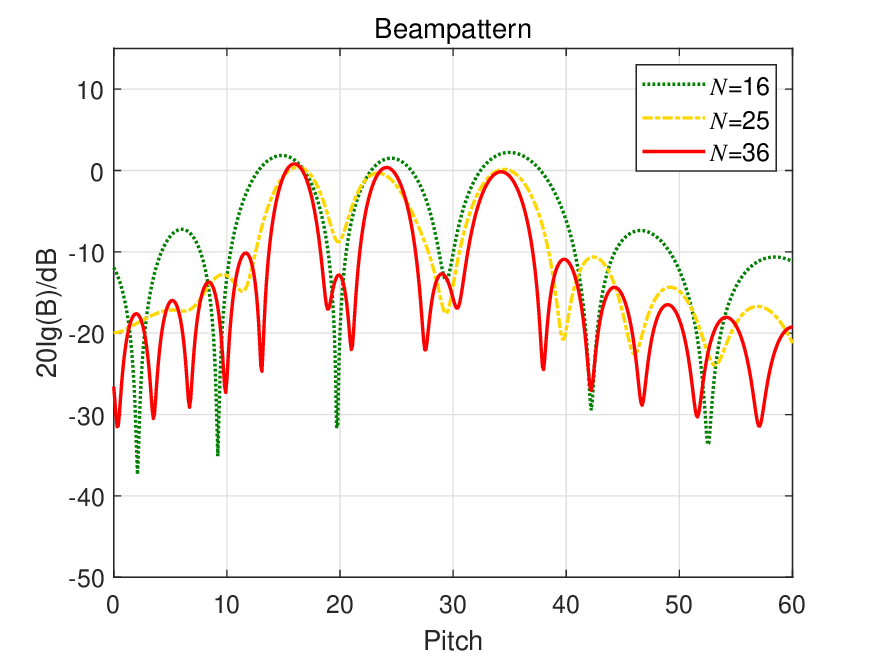}\\
			(a) Azimuth & (b) Pitch\\
		\end{tabular}
	\end{minipage}
	\caption{Beampattern under different TRIS elements ($R_{th}=1Mbits$).}
	\label{Beampattern}
\end{figure}

Finally, in order to visualize which angles of the beam are enhanced, the beampatterns in azimuth and pitch are plotted in Fig. \ref{Beampattern}. The beampatterns show that azimuths of $47.72\degree$, $35.53\degree$, and $63.43\degree$ degrees are enhanced and pitch angles of $33.92\degree$, $16.20\degree$, and $24.09\degree$ degrees are enhanced, which are close to the angle calculated at the start coordinates of the terminals. It is worth noting that as the number of TRIS elements increases, the beamwidth narrows at specific angles in the beampatterns.
\section{Conclusions}

In this paper, we proposed a TRIS-empowered architecture for ISAC networks. For scenario-specific applications, we provided the QoSs of communication and sensing. To tackle the communication and sensing interference, we utilized the RSMA to manage the interference. Moreover, the multi-stream communication and sensing can be simultaneously achieved due to the characteristic of the TRIS. Based on these QoSs, a joint communication and sensing design algorithm was proposed. Numerical simulations verified the effectiveness of the proposed architecture and a trade-off can be clearly seen between the communication and sensing.  

\appendices
\section{Proof of Lemma 1}
In this section, we compute the LRT detector for TRIS empowered ISAC network. As above-mentioned, the optimal detector under the Neyman-Pearson sense is LRT, which is given by \cite{Trees2001}

\begin{equation}
		\setlength{\abovedisplayskip}{3pt}
	\setlength{\belowdisplayskip}{3pt}
	T = \log \frac{{f\left( {r\left( t \right)|{{\cal H}_1}} \right)}}{{f\left( {r\left( t \right)|{{\cal H}_0}} \right)}}\mathop \gtrless \limits_{{H}_0}^{{H}_1}  \delta.
	\label{TH}
\end{equation}
where ${f\left( {r\left( t \right)|{{\cal H}_0}} \right)}$ and ${f\left( {r\left( t \right)|{{\cal H}_1}} \right)}$ are the probability density function (PDF) of the observation given the null and alternative hypotheses, respectively.
 
 Firstly, we derive the PDF of the received measurements under both the null and alternative hypotheses. Consider the alternative hypothesis in (\ref{h1}), where ${O_1} = {\bf{c}}\left( {{\theta _1},{\varphi _1}} \right){{\bf{p}}_c},$ and ${\hat x_1} = \int {r\left( t \right)s_c^ * \left( {t - {\tau _{d,1}}} \right){e^{ - j2\pi {f_{d,1}}t}}dt}.$ $\left( a \right)$ holds because of $\int {{s_{c,i}}\left( t \right){s_{c,j}}\left( t \right)} dt = {\delta _{ij}}$, where ${s_c} = {s_{c,1}} +  \cdots  + {s_{c,K}}$.

Then, the PDF of the received measurements under the null hypothesis is given by
\setcounter{equation}{67} 
\begin{equation}
		\setlength{\abovedisplayskip}{3pt}
	\setlength{\belowdisplayskip}{3pt}
	f\left( {r\left( t \right)|{{\cal H}_0}} \right)=c\exp \left( { - \frac{1}{{\sigma _r^2}}\int {{{\left| {r\left( t \right)} \right|}^2}dt} } \right).
	\label{h0}
\end{equation}

Finally, combining (\ref{TH}), (\ref{h1}), and (\ref{h0}) results in the following detector

\begin{equation}
		\setlength{\abovedisplayskip}{3pt}
	\setlength{\belowdisplayskip}{3pt}
	T = \log f\left( {r\left( t \right)|{{\cal H}_1}} \right) - \log f\left( {r\left( t \right)|{{\cal H}_0}} \right) = {\left| {{{\hat x}_1}} \right|^2}\mathop \gtrless \limits_{{H}_0}^{{H}_1}  \delta.
\end{equation}
$\hfill\blacksquare$

\section{FIM Derivation}

The FIM derivation of the parameters ${{\bf{\xi }}_2} = {\left[ {{\tau _{d,2}},{f_{d,2}},{\theta _2},{\varphi _2}} \right]^T}$ can be calculated by

\begin{equation}
	{\left[ {{{\bf{I}}_2}} \right]_{i,j}} = \frac{2}{{\sigma _r^2}}{\mathop{\rm Re}\nolimits} \left\{ {\sum\limits_{l = 1}^L {\frac{{\partial {L^H_2}{{\left[ l \right]}}}}{{\partial {\xi _i}}}\frac{{\partial {L_2}\left[ l \right]}}{{\partial {\xi _j}}}} } \right\},{\rm{ }}i,j \in \left\{ {1,2,3,4} \right\}.
\end{equation}
where the partial derivatives with respect to ${\theta _2}$ and ${\varphi _2}$ can be easily calculated by the above equation. However, due to non differentiability of time delay $\tau _{d,2}$ calculation, we transform it as follow\cite{Kay1995}

\begin{equation}
\begin{aligned}
	{I_{{\tau _{d,2}}{\tau _{d,2}}}} &= \frac{2}{{\sigma _r^2}}{\mathop{\rm Re}\nolimits} \left\{ {{{\sum\limits_{l = 1}^L {\left( {\frac{{\partial {L_2}\left[ {l;{\tau _{d,2}}} \right]}}{{\partial {\tau _{d,2}}}}} \right)} }^2}} \right\}\\
	&= \frac{2}{{\sigma _r^2}}{\mathop{\rm Re}\nolimits} \left\{ {\sum\limits_{l = {l_0} + 1}^{{l_0} + M} {{{\left( {\frac{{\partial {L_2}\left[ {l\Delta  - {\tau _{d,2}}} \right]}}{{\partial {\tau _{d,2}}}}} \right)}^2}} } \right\}\\
	& = \frac{2}{{\sigma _r^2}}{\mathop{\rm Re}\nolimits} \left\{ {\sum\limits_{l = {l_0} + 1}^{{l_0} + M} {{{\left( {{{\left. {\frac{{\partial {L_2}\left( t \right)}}{{\partial t}}} \right|}_{t = l\Delta  - {\tau _{d,2}}}}} \right)}^2}} } \right\}\\
	& = \frac{2}{{\sigma _r^2}}{\mathop{\rm Re}\nolimits} \left\{ {\sum\limits_{l = l1}^M {{{\left( {{{\left. {\frac{{\partial {L_2}\left( t \right)}}{{\partial t}}} \right|}_{t = l\Delta }}} \right)}^2}} } \right\}\\
	& = \frac{2}{{\sigma _r^2}}{\mathop{\rm Re}\nolimits} \left\{ {\frac{1}{\Delta }\int_0^T {{{\left( {\frac{{\partial {L_2}\left( t \right)}}{{\partial t}}} \right)}^2}dt} } \right\}\\
	& = \frac{{2A{{\left| {{\alpha _2}} \right|}^2}}}{\Delta{\sigma _r^2}}{\mathop{\rm Re}\nolimits} \left\{ {\int_0^T {\left[ {{{\left( {\frac{{\partial s\left( t \right)}}{{\partial t}}} \right)}^2} + {{\left( {4\pi {f_{d,2}}WT} \right)}^2}} \right]dt} } \right\},
\end{aligned}
\end{equation}
where $\int_0^T {{{\left( {\frac{{\partial s\left( t \right)}}{{\partial t}}} \right)}^2}dt}  = \varepsilon \overline {{F^2}},$ $\varepsilon  = \int_0^T {{s^2}\left( t \right)dt},$ $\frac{1}{\Delta }=2W$, and $\overline {{F^2}}  = \frac{{\int_0^T {{{\left( {\frac{{\partial s\left( t \right)}}{{\partial t}}} \right)}^2}dt} }}{{\int_0^T {{s^2}\left( t \right)dt} }}$. Assume the signal is a Gaussian pulse and that $s(t)$ is nonzero over the interval $[0,T]$, and then $\overline {{F^2}}  = {{\sigma _F^2} \mathord{\left/
		{\vphantom {{\sigma _F^2} 2}} \right.
		\kern-\nulldelimiterspace} 2}$. Thus, the FIM of the time delay can be expressed as
\begin{equation}
	{I_{{\tau _{d,2}}{\tau _{d,2}}}} = \frac{{4WA{{\left| {{\alpha _2}} \right|}^2}}}{{\sigma _r^2}}\left( {\varepsilon \overline {{F^2}}  +8WT^3 {{\left( {\pi {f_{d,2}}} \right)}^2}} \right).
\end{equation}
Notice that there has $\int_0^T {s\left( t \right)} \frac{{\partial s\left( t \right)}}{{\partial t}}dt = \left. {\frac{{{s^2}\left( t \right)}}{2}} \right|_0^T = 0$, which is due to the fact that $s(t)$ is smooth in $[0,T]$, we may assume $s(0) = s(T)$. This feature can be used to compute other delay-dependent FIMs, which are ignored here in this paper.

$\hfill\blacksquare$
\newcounter{TempEqCnt1}                         
\setcounter{TempEqCnt1}{\value{equation}} 
\setcounter{equation}{66}             
\begin{figure*} 
	\centering
	\begin{equation}
		\begin{array}{l}
			f\left( {r\left( t \right)|{{\cal H}_1}} \right) = \int {f\left( {r\left( t \right)|{{\cal H}_1},{\bf{\alpha }}} \right)} f\left( {\bf{\alpha }} \right)d{\bf{\alpha }}\\
			= \int {c\exp \left( { - \frac{{\rm{1}}}{{\sigma _r^2}}\int {{{\left| {r\left( t \right) - {\alpha _1}{\bf{c}}\left( {{\theta _1},{\varphi _1}} \right){{\bf{p}}_c}{s_c}\left( {t - {\tau _{d,1}}} \right){e^{j2\pi {f_{d,1}}t}}} \right|}^2} - {{\left| {{\alpha _1}} \right|}^2}} } \right)} d{\alpha _1}\\
			\mathop  = \limits^{(a)} c\exp \left( { - \frac{1}{{\sigma _r^2}}\int {{{\left| {r\left( t \right)} \right|}^2}dt} } \right)\int {\exp \left( { - \frac{1}{{\sigma _r^2}}\left( { - 2{\mathop{\rm Re}\nolimits} \left\{ {{\alpha _1}{O_1}{{\hat x}_1}} \right\} + {{\left| {{\alpha _1}} \right|}^2}{{\left| {{O_1}} \right|}^2} + \sigma _r^2{{\left| {{\alpha _1}} \right|}^2}} \right)} \right)} d{\alpha _1}\\
			= cf\left( {r\left( t \right)|{{\cal H}_0}} \right)\exp \left( {\frac{{{{\left| {{O_k}{{\hat x}_1}} \right|}^2}}}{{\sigma _r^2\left( {{{\left| {{O_1}} \right|}^2} + \sigma _r^2} \right)}}} \right)\int {\exp \left( { - \frac{1}{{\sigma _r^2}}{{\left| {\sqrt {{{\left| {{O_1}} \right|}^2} + \sigma _r^2} {\alpha _1} - \frac{{\left| {{O_1}{{\hat x}_1}} \right|}}{{\sqrt {{{\left| {{O_1}} \right|}^2} + \sigma _r^2} }}} \right|}^2}} \right)} d{\alpha _1}\\
			= c'f\left( {r\left( t \right)|{{\cal H}_0}} \right)\exp \left( {\frac{{{{\left| {{{\hat x}_1}} \right|}^2}}}{{\sigma _r^2\left( {1 + {{\sigma _r^2} \mathord{\left/
									{\vphantom {{\sigma _r^2} {{{\left| {{O_1}} \right|}^2}}}} \right.
									\kern-\nulldelimiterspace} {{{\left| {{O_1}} \right|}^2}}}} \right)}}} \right).
		\end{array}
		\label{h1}
	\end{equation}
	\hrulefill
\end{figure*}
\setcounter{equation}{\value{TempEqCnt1}} 
\bibliographystyle{IEEEtran}
\bibliography{IEEEabrv,draft}

%
%
%
%
%
%
%
%

\end{document}